\documentclass[twocolumn]{aastex63}
\usepackage{amsmath}

\usepackage[FIGTOPCAP]{subfigure}
\usepackage [english]{babel}
\usepackage [autostyle, english = american]{csquotes}
\MakeOuterQuote{"}

\received{****}
\revised{*****}
\accepted{*****}

\shorttitle{Merger Signatures in Galaxies}
\shortauthors{Lambrides et al.}
\graphicspath{{./}{figures/}}

\begin{document}

\title{Merger or Not: Accounting for Human Biases in Identifying Galactic Merger Signatures}

\correspondingauthor{Erini Lambrides}
\email{erini.lambrides@jhu.edu}

\author[0000-0002-0786-7307]{Erini L. Lambrides}
\affil{Department of Physics and Astronomy, Johns Hopkins University,Bloomberg Center, 3400 N. Charles St., Baltimore, MD 21218, USA}

\author[0000-0002-5437-6121]{Duncan J. Watts}
\affil{Institute of Theoretical Astrophysics, University of Oslo, P.O. Box 1029 Blindern,N-0315 Oslo, Norway}
\affil{Department of Physics and Astronomy, Johns Hopkins University,Bloomberg Center, 3400 N. Charles St., Baltimore, MD 21218, USA}

\author{Marco Chiaberge}
\affil{AURA for the European Space Agency (ESA), ESA Office, Space Telescope Science Institute, 3700 San Martin Drive, Baltimore, MD 21218, USA}
\affil{Department of Physics and Astronomy, Johns Hopkins University,Bloomberg Center, 3400 N. Charles St., Baltimore, MD 21218, USA}

\author{Kirill Tchernyshyov}
\affil{Department of Astronomy, University of Washington, Seattle, WA, USA}

\author{Allison Kirkpatrick}
\affil{Department of Physics and Astronomy, University of Kansas, Lawrence, KS 66045, USA}

\author{Eileen T. Meyer}
\affil{Department of Physics, University of Maryland, Baltimore County, 1000 Hilltop Circle, Baltimore, MD 21250, USA}

\author{Timothy Heckman}
\affil{Department of Physics and Astronomy, Johns Hopkins University,Bloomberg Center, 3400 N. Charles St., Baltimore, MD 21218, USA}

\author{Raymond Simons}
\affil{Space Telescope Science Institute, 3700 San Martin Drive Baltimore, MD 21218, USA}

\author{Oz Amram}
\affil{Department of Physics and Astronomy, Johns Hopkins University,Bloomberg Center, 3400 N. Charles St., Baltimore, MD 21218, USA}

\author{Kirsten R. Hall}
\affil{Schmidt Science Fellow}
\affil{Atomic and Molecular Physics Division, Harvard-Smithsonian Center for Astrophysics, 60 Garden Street, Cambridge, MA, 02138, USA}

\author{Arianna Long}
\affil{Department of Physics and Astronomy, University of California, Irvine, CA 92697, USA}

\author{Colin Norman}
\affil{Space Telescope Science Institute, 3700 San Martin Drive Baltimore, MD 21218, USA}
\affil{Department of Physics and Astronomy, Johns Hopkins University,Bloomberg Center, 3400 N. Charles St., Baltimore, MD 21218, USA}



\begin{abstract}

Significant galaxy mergers throughout cosmic time play a fundamental role in theories of galaxy evolution. The widespread usage of human classifiers to visually assess whether galaxies are in merging systems remains a fundamental component of many morphology studies. Studies that employ human classifiers usually construct a control sample, and rely on the assumption that the bias introduced by using humans will be evenly applied to all samples. In this work, we test this assumption and develop methods to correct for it. Using the standard binomial statistical methods employed in many morphology studies, we find that the merger fraction, error, and the significance of the difference between two samples are dependent on the intrinsic merger fraction of any given sample. We propose a method of quantifying merger biases of individual human classifiers and incorporate these biases into a full probabilistic model to determine the merger fraction and the probability of an individual galaxy being in a merger. Using 14 simulated human responses and accuracies, we are able to correctly label a galaxy as ''merger'' or ''isolated'' to within 1\% of the truth. Using 14 real human responses on a set of realistic mock galaxy simulation snapshots our model is able to recover the pre-coalesced merger fraction to within 10\%. Our method can not only increase the accuracy of studies probing the merger state of galaxies at cosmic noon, but also can be used to construct more accurate training sets in machine learning studies that use human classified data-sets.

\end{abstract}

\keywords{statistics -- methods --galaxies-- mergers}


\section{Introduction} \label{sec:intro}

There is mounting observational and theoretical evidence that significant galactic mergers, where one galaxy is at least the tenth of the mass of the other, are an important component of galaxy evolution models which aim to explain the size, shape, and mass distributions of galaxies in the Universe \citep[see][for a review]{cons_review}. Observational estimates of the rates of significant galaxy mergers have not converged for a variety of merger types. Even studies of the same observational field, with similar wavelength coverage, can yield disparate merger rate estimates \citep[][]{mantha2018,duncan19}. Thus, robustly and consistently identifying systems that are ongoing (galaxy pairs or pre-coalescence) or recently have undergone a significant merger (near- or post-coalescence) is important. At higher redshifts, merger identification can become increasingly difficult due to the potential for faint merger signatures to be undetectable \citep{lotz2004}.  

For example, some hydrodynamic simulations of galaxy mergers predict that as the galaxies coalesce, gravitational forces funnel gas toward the center, which provides a fuel reservoir to feed the central super-massive black hole and to form large numbers of stars in a nuclear starburst \citep{dimatteo05}. Between redshifts 1.5 and 2.5, activity of growing central super-massive black-holes (herein referred to as active galactic nuclei -- AGN) and star-formation (SF) activity appear to peak \citep{madau}. Galaxy mergers with comparable mass ratios (i.e., major mergers) are one of the most popular mechanisms invoked to explain the similar evolution of the AGN activity and SF rates during this cosmic epoch \citep{cons_review}. Some results are in tension with this picture. For example, empirical and theoretical studies find a connection between mergers and local ultra-luminous infrared galaxies \citep{sanders96,veilleux09}, local AGN \citep{koss10,ellison13,ellison19}, and high-luminosity AGN \citep{urrutia08,treister12,glikman15,donley18}. In contrast, ample research finds no connection between mergers and X-ray detected AGN \citep{gabor09,georgakakis09,kocev12}, high-luminosity AGN \citep{villforth14,villforth17,marian19}, and low-to-intermediate luminosity AGN \citep{grogin05,schawinski11,rosario15}. 

Selection effects introduced through the construction of the AGN sample may play a role in explaining some of the disparate conclusions between AGN morphology studies. For example, dust obscuration may play a significant role in the observed (or lack of) connection between AGN and mergers. The merger fraction is higher for samples of infra-red (IR) selected AGN versus X-ray selected AGN perhaps due to the effect of dust-attenuation \citep{veilleux09,koss10,kocev15}. Though, studies of sources with similar AGN selection criteria still yield conflicting merger fractions. For example the merger enhancement of X-ray selected heavily-obscured AGN at both higher and lower redshifts yield conflicting results \citep[i.e.][]{schawinski12,kocev15,koss16,lanzuisi2018,li2020}. An ill-studied reason for this disagreement may be the diverse array of merger detection methods and/or statistical methods used to characterize the statistical significance of the results within each study.  

The variety of merger detection methods used to assess the morphology of galaxies can be broadly placed in two regimes: qualitative and automated. Qualitative methods rely on an observer or group of observers who classify each image by eye. Automated methods employ a pixel by pixel analysis of the image to identify the morphological class of the galaxy. Some automated methods require highly spectroscopic complete observations, like the close pairs method, which uses redshift and on-sky distances to identify pairs of galaxies that are within some distance threshold. Non-parametric automated methods, such as the second-order moment of the brightest 20\% of light, the Gin$i$ coefficient, and the CAS parameters (concentration, asymmetry, clumpiness) use pixel based algorithms to detect asymmetries, double nuclei, tidal tails and/or other disturbances \citep[for examples see][]{abraham96,conselice2000,lotz2004}. As shown in \citet{huertas2015}, some of these  methods can have mis-classification rates as high as 20\%, and each suffers from biases where certain merging systems are preferentially identified. 

Automated methods that employ deep learning techniques, a sub-field of machine learning based on artificial neural networks with representation learning, to classify galaxy morphology are promising due to their ability to classify quickly and their model independence \citep[for example][]{dielman15,ackermann2018,pearson2019,deepmerge20}. In particular, a variety of deep learning merger morphology studies train their algorithms on data-sets that have been visually classified by humans or test the accuracy of their schema compared to visually classified ''truth'' data-sets. Many of these recent deep-learning schema are trained off of the \textit{Galaxy Zoo} catalogue of classifications of galaxies from the Sloan Digital Sky Survey (SDSS) \citep{galaxyzoo1,galaxyzoo2}. Most of these ML implementations morphologically analyse galaxy samples at moderate to low redshifts. For example, \citet{pearson2019}, employed a deep learning algorithm that was trained not only on visually classified objects via \textit{Galaxy Zoo}, but also mock images with known truths from the Eagle Simulations. When applying a convolutional neural network on the SDSS images, an accuracy of 91.5\% was achieved. When passing the simulated EAGLE images through the SDSS trained neural network, the accuracy drops to 64.6\%. The \citet{pearson2019} framework uses SDSS galaxies with redshifts less than 0.1, and simulated EAGLE galaxies with redshifts less than 1.0. As is noted in \citet{pearson2019}, due to the potential redshift evolution of general galaxy properties, such as gas and dust content, a network trained on low-redshift galaxies is not expected to be reliable for higher redshift galaxies. 

Furthermore, any deep learning model trained on human classifications will carry any bias that still persists in the human classified training set. Despite the great potential of these classes of algorithms for automated merger identification, there currently is not a robust enough tool to handle the diverse presentations of merging galaxies, particularly at higher redshifts. Thus, visual human classification is still a method that is commonly employed in the literature to identify moderately large samples of merging galaxies at z $> 1.0$. 

Image-based morphology studies of galaxies at higher-redshifts are difficult. Beyond $z\sim 1$, optical imaging surveys begin to probe the rest-frame UV morphologies of galaxies. This is useful for probing the most active regions of un-obscured star-formation, but may miss obscured gaseous and stellar features associated with merging systems (e.g., dusty tidal tails, dusty shells, and large-scale dust and gas asymmetries). When using humans as classifiers there are a variety of assumed biases most studies try to take into account. It is inevitable that any given classifier will show a particular bias. For example, some observers may be more inclined to classify objects as mergers even if the objects display minor disturbances unrelated to galaxy encounters. The most common way of accounting for human classifier bias is to construct a control sample.  The classifiers assess the morphology of the control sample, and report merger fractions of their galaxy population of interest in the context of their relative differences between the control sample. In addition to constructing a control sample, some studies try to maximize the number of individual human classifiers. For projects like \textit{Galaxy Zoo}, there is an average of 39 classifiers per object, and they report merger classifications on a per galaxy basis. 

When comparing merger fractions of a population of objects to a control sample, careful analysis of the error bars is critical in order to determine if  a significant difference exists between the population of interest and the control sample. The variety of statistical treatments used in reporting merger fractions from human classified datasets makes comparisons between studies difficult. For example, some studies assume a binomial distribution to model the number of mergers from aggregate classifications given by a group of human classifiers \citep[i.e.][]{ellison13,kocev15,ellison19}. Other studies have employed rank-choice voting and model the \textit{probability} of the number of mergers using a beta distribution \citep[i.e.][]{mechtley16,marian19,marian20}. All of the above studies compare the significant of their merger fraction against a similar statistically analysed control sample, with the assumption that the human classification bias is evenly applied amongst samples.  

In this work, we test the critical assumption that the bias present in human classification is evenly applied to both the population and control datasets. In \autoref{sec:problem}, we find it is not, and that the effect of human bias is a function of the intrinsic merger fraction of the sample being classified. In \autoref{sec:merger_frac_like}, we propose a self-consistent statistical framework to use estimates of an individual human classifier's accuracy to derive a data-driven merger fraction. In \autoref{sec:per_gal}, we describe how we can use the data-driven merger fraction and human classifier accuracy to yield merger assessments on a per-galaxy basis. In \autoref{sec:conclusions}, we discuss the implications and applications of our statistical framework.

\section{Idealized Problem and Issues with the Conventional Approach} \label{sec:problem}

The fundamental setup for a morphology study is as follows;  given a set of $n$ galaxies and $N$ independent classifications of each galaxy, what is the estimated merger fraction and error on the estimate for the given population? 
Most studies treat this as a binomial process with two outcomes: ''merger'' and ''not merger'', where the fraction of galaxies in a merger is given by $f_M$.

Generally what is reported is the merger fraction of the science sample, the merger fraction of the control sample, and the difference between the two. For example, suppose three classifiers assess 50 galaxies in two sets of samples, and report $\{30,33,31\}$ mergers in the science sample, and $\{10,13,15\}$ mergers in the control sample. Conventionally, the estimate of the merger fraction for each sample would be the mean of the individual measured merger fractions, and the error would be determined using binomial statistics. The significance of the estimated merger fraction in the science sample case is determined using a differential approach. In the above example, the mean merger fraction of the science sample is $\sim2.5$ times greater than the control sample, and thus some significance of the estimated merger rate of the science sample would be assumed. Part of why most merger studies report results using differential or relative treatments is because the unknown biases of a classifier's measurements is assumed to be applied evenly to both samples and thus should cancel out. 

\begin{figure}
    \centering
    \includegraphics[width=\columnwidth]{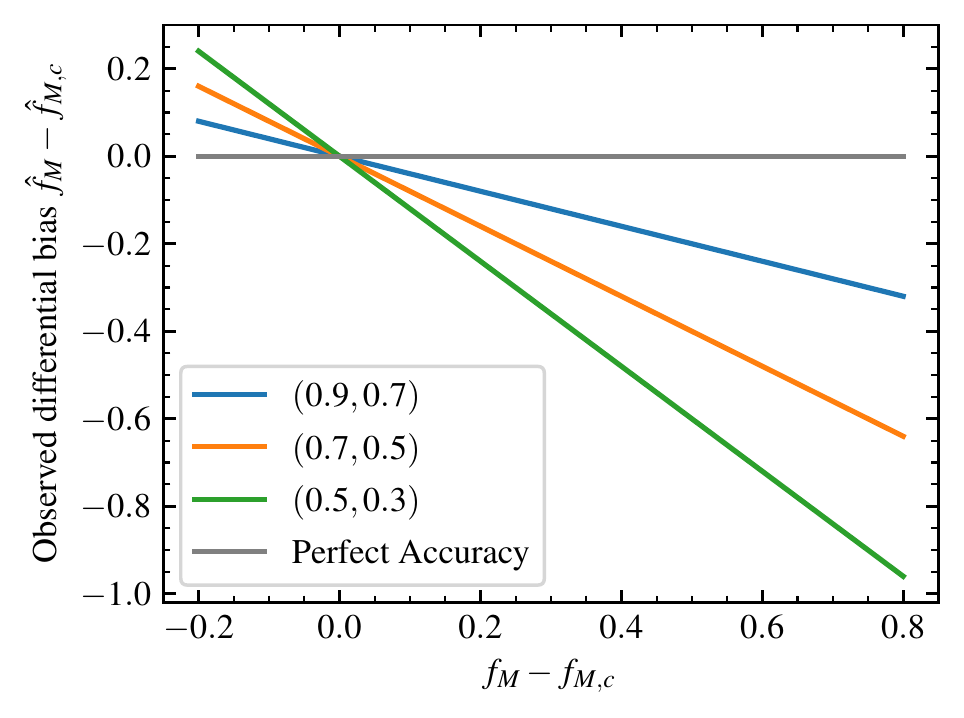}
    \caption{Observed differential merger fraction bias as a function of true differential merger fraction: Using \autoref{eq:bias}, the difference between a control sample and a science sample gives a result that depends both on the intrinsic differential merger fraction and the accuracy of the classifier.
    }
    \label{fig:merger_diff_binom}
\end{figure}

Closer examination shows that this framework is internally inconsistent. This treatment assumes that we are showing 3 separate samples to each person, but in fact they are looking at the same galaxy and disagreeing. If for example, in the control sample each classifier identified a similar number of mergers, but they disagreed with each other on the classification of individual objects, the statistical framework would not encapsulate important sources of error. 
At a fundamental level, if there is a disagreement amongst classifiers on a given classification: due to the binomial nature of the experimental set-up, one set of classifiers will be incorrect. To formalize this, we can say that if someone is shown a merging (isolated) galaxy, they classify it correctly with probability $r_M$ ($r_I$). Therefore, if somebody is shown $N_M$ mergers and $N_I$ isolated galaxies, on average they will report
$\hat N_M=r_M N_M + (1-r_I)N_I$ mergers. The inclusion of the $(1-r_I)N_I$ term represents the amount of galaxies that were incorrectly classified as isolated and are truly mergers.   

The use of relative significance between comparing the merger fractions of the science and the control sample does not remove this issue. In the control sample, the classifier will report $\hat N_{M,c}=r_MN_{M,c}+(1-r_I)N_{I,c}$ mergers on average, meaning the difference between the merger fractions depends on both the accuracy of an individual classifier and the intrinsic merger fraction of the sample. Thus by re-writing $\hat N_{M}$ and $\hat N_{M,c}$ in terms of the merger fraction for each sample and taking the difference: 
\begin{align}
    \langle \hat f_M\rangle
    &=r_Mf_M+(1-r_I)(1-f_M)
    \label{eq:fm_hat}
    \\
    \langle \hat f_{M,c}\rangle &=
    r_Mf_{M,c}+(1-r_I)(1-f_{M,c})
    \\
    \langle \Delta\hat f_M\rangle &= 
    \langle \hat f_M\rangle - \langle \hat f_{M,c}\rangle 
    \\
    \langle \Delta\hat f_M\rangle  &=
    r_M\Delta f_M
    -(1-r_I)\Delta f_M
    \nonumber
    \\
    &=\Delta f_M[r_M+r_I-1]
\end{align}
we find the difference between the merger fractions of the two samples is still dependent on the intrinsic merger fraction of each sample. Equation (3) can then be used to quantify the systematic error due to human classification in the difference between the merger fractions as
\begin{equation} \label{eq:bias}
b=\Delta f_M[r_M+r_I-2].
\end{equation}

The only time when the bias would be equal to zero is if the intrinsic merger fraction of the two samples were identical. This is highly significant, because most morphology studies test whether there is a difference between the science sample and control sample. In \autoref{fig:merger_diff_binom}, we show three examples of this effect. Using the previous example of three classifiers assessing 50 galaxies for two sets of samples, we calculate the bias as parametrized in \autoref{eq:bias} as a function of the difference of the intrinsic merger fractions for each sample. We calculate this function in four different test cases of mean observer accuracy. The blue line represents a class of observers that are very accurate in measuring merging systems and slightly less accurate at measuring isolated systems. The orange line is for a class of observers who are slightly less accurate at identifying merging galaxies and isolated galaxies. The green line represents a class of observers whose accuracy is poor for both merging and isolated systems, and the black line for classifiers with perfect accuracy. We see in all three classes of observers with non-perfect accuracy the degree of systematic bias from the truth changes as a function of the intrinsic merger fraction of each sample. Thus, if a hypothetical study finds a difference in the estimated merger fractions of their science sample and control samples, assuming the accuracy of their classifiers is not taken into account, disentangling whether the difference is due to real or simply systematic error is impossible.   


\begin{figure}
    \centering
    \includegraphics[width=\columnwidth]{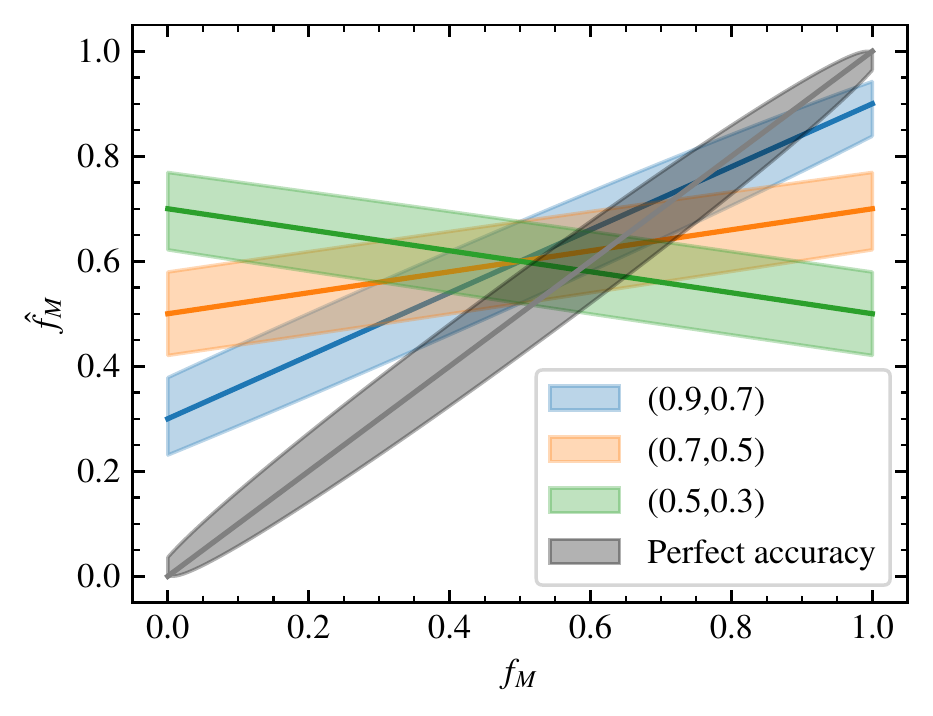}
    \caption{Intrinsic Merger Fraction vs Measured Merger Fraction: The black line corresponds to human classifiers with perfect accuracy. The blue, green, and orange lines correspond to different merger,isolated accuracy pairs. The shaded regions correspond to 68\% confidence levels governed by the beta distribution.}
    \label{fig:beta_dist}
\end{figure}

\begin{figure}
    \centering
    \includegraphics[width=\columnwidth]{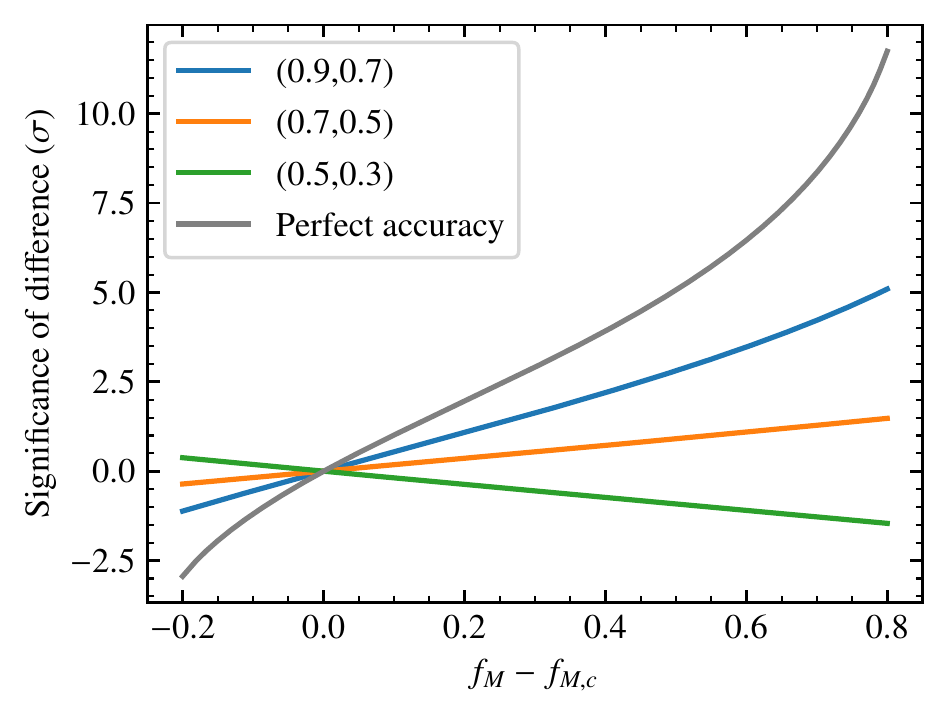}
    \caption{Intrinsic Difference of Merger Fractions vs Significance of the  Measured Difference Between Merger Fractions: The colors correspond to the same definitions as of \autoref{fig:beta_dist}. We test how the significance of the measured difference between a population sample and a control sample changes as a function of the intrinsic population sample merger fraction. We use the same intrinsic control sample merger fraction, $f_{M,c} = 0.2$, for each difference.}
    \label{fig:significance}
\end{figure}

We next explore how the unequal effect of this bias hinders meaningful statistical interpretation of sample difference measures between two merger fractions. Using the standard binomial statistics approach \citep[as seen in][]{chiaberge15,villforth17}, one would use a proportion test to calculate confidence intervals of a given merger fraction and then use a hypothesis test to calculate the probability or significance of a difference between two samples given the null. In \autoref{fig:beta_dist}, we show the effect of inaccurate classifiers on the recovered difference on a simulated sample of 50 galaxies. As shown in \autoref{eq:fm_hat}, $\hat f_M$ is a function of the true $f_M$ and the accuracy of the classifier. Using the beta-distribution 68\% confidence levels, for each $r_M$,$r_I$ pair shown, there are regions in the parameter space with many standard deviations of difference between the $\hat f_M$ that would be reported by a perfect observer (black line) and by an inaccurate observer.

As previously mentioned, often these errors are mitigated by estimating the merger fraction of a control sample with an unknown, but likely lower, merger fraction than the science sample in question, and the relative enhancement is reported. Assuming that $r_M$ and $r_I$ are independent of the class of object being classified, we can estimate the significance of the difference between a control sample's merger fraction, here $f_{M,c}=0.2$, to see the size of the effect. We derive for each case the estimate of the control sample's merger fraction, $\hat f_{Mc}=r_M f_{Mc}+(1-r_I)(1-f_{Mc})$ and its uncertainty $\sigma_{f_{Mc}}$. To quantify the difference between a control sample and science sample, we plot $(\hat f_M-\hat f_{Mc})/\sigma$, where $\sigma^2=\sigma_{f_{Mc}}^2 + \sigma_{f_M}^2$ using the errors derived from the proportion test. As shown in \autoref{fig:significance}, the general effect of this is to reduce the size of the measured difference. This does not imply that all previous merger studies have reported a lower significance than the actual truth, but rather if human classification bias is not constrained or accounted for and a null-significance is reported it is difficult to to deduce whether the null result is intrinsically true or a human classifier accuracy effect. We note this reduction of significance is difficult to infer the validity of previous merger studies, due to the specific parametrization of accuracy and statistical tests used in this section. We do stress, the significance of the effect has a clear dependence on the accuracy of the classifier, and this effect is not mitigated by performing a comparison between the $\hat f_M$ of the two samples using simple binomial statistical approaches.


\section{A Bayesian Upgrade to the Frequentist Approach: The Number of Mergers Likelihood} \label{sec:merger_frac_like}


\begin{deluxetable*}{ll}
\tablecaption{Variable definitions for the Merger Fraction Likelihood}
\tablehead{\colhead{Symbol} & \colhead{Definition}}

\startdata
$N_{X}$ & Number of objects of type $X$ in the sample\\
$f_M$ & Merger fraction of sample $f_M\equiv N_M/(N_M+N_I)$\\
$N_{X,\mathrm{syn}}$ & Number of mock  objects of type $X$ shown to a classifier\\
$r_{X,i}$ &  Probability of classifier $i$ identifying object type $X$ correctly\\
$\hat N_{X,\mathrm{syn},i}$ & Number of mock  objects of type $X$ correctly identified by classifier $i$\\
$\hat N_{X,1,i}$ & Number of objects \textit{correctly} identified as type $X$ by classifier $i$\\
$\hat N_{X,2,i}$ & Number of objects \textit{incorrectly} identified as type $X$ by classifier $i$\\
$\hat N_{X,i}$ & Number of objects identified as type $X$ by classifier $i$, $\hat N_{X,i}\equiv \hat N_{X,1,i}+\hat N_{X,2,i}$\\
$\hat f_M$ & Estimated merger fraction given the set of all $\{\hat N_{M,i},\hat N_{I,i},\hat N_{M,\mathrm{syn},i},\hat N_{I,\mathrm{syn},i}\}$.
\enddata
\tablecomments{We do not use the incorrectly classified mock  galaxies in our likelihood, since we know the ground truth and do not need to marginalize over this parameter.}
\end{deluxetable*}

As discussed in \autoref{sec:intro}, it is difficult to accurately characterize whether a galaxy is undergoing a merger or is isolated. Because of this, it is inevitable that any given classifier will obtain a merger fraction that is different than another's. Some previous works have assumed that this bias is similar for the data and the control sample, and test their results against the null hypothesis that the intrinsic fractions are identical. As shown in \autoref{fig:significance}, if the underlying merger fraction of the two populations are significantly different, the significance of the result will be affected by this bias. In this section, we present a method that is built upon on the standard binomial approach of determining the number of mergers in a sample while taking into account the effect of human inaccuracy. 

In order to estimate the true underlying merger fraction, we can estimate the bias in an individual classifier's assessment on a sample with a known intrinsic merger fraction, and then optimally combine the individual classifier uncertainties on the sample where the intrinsic merger fraction is unknown. We perform this analysis assuming that there are two binomial processes for each classifier; (1) the probability of classifying galaxies accurately as mergers, and (2) the probability of inaccurately classifying isolated galaxies as mergers.

\subsection{The Merger Fraction Likelihood}

\begin{figure*}
    \centering
    \includegraphics[width=0.7\textwidth]{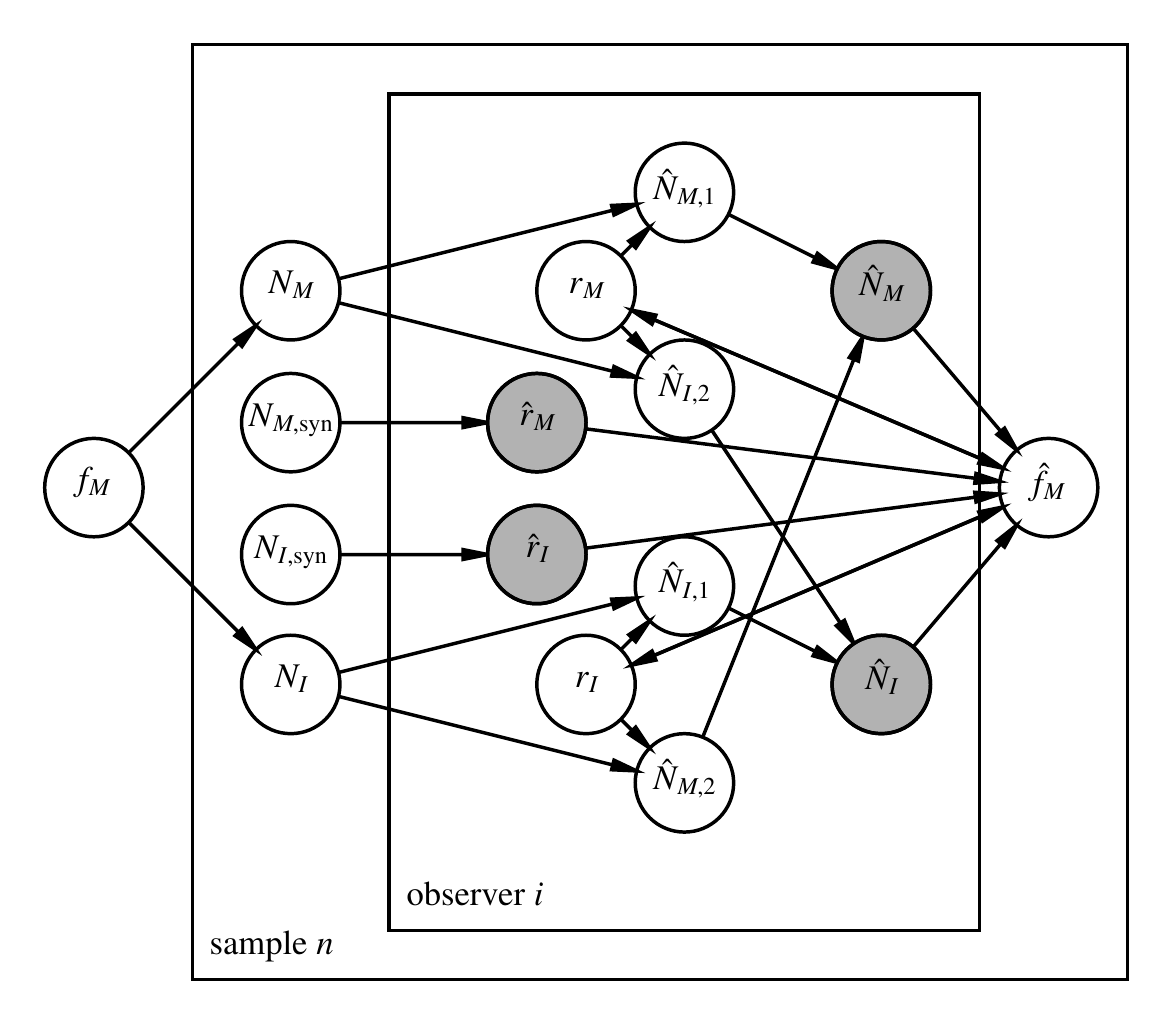}
    \caption{Graphical model of obtaining the merger fraction using the likelihood defined in 
    \autoref{eq:posterior}.}
    \label{fig:stat_schema}
\end{figure*}

The total number of claimed mergers is $\hat{N}_{M} = \hat{N}_{M,1} + \hat{N}_{M,2}$. Given $N_{M}$ mergers in a sample, the probability of a classifier correctly measuring $\hat{N}_{M,1}$ mergers in a given sample is
\begin{equation}
P(\hat{N}_{M,1}\mid r_{M}, N_{M}) = \binom{N_{M}}{\hat{N}_{M,1}} r_{M}^{\hat{N}_{M,1}}(1-r_{M})^{N_{M} - \hat{N}_{M,1}}.
\end{equation}
At the same time, if we have $N_I$ isolated galaxies, the classifier will incorrectly classify an isolated galaxy as a merger with probability $1-r_I$. We define the number of isolated galaxies incorrectly identified as mergers as $\hat N_{M,2}$, which follows the probability distribution
\begin{equation}
P(\hat{N}_{M,2}\mid r_{I}, N_{I}) = \binom{N_{I}}{N_{I} - \hat{N}_{M,2}} r_{I}^{N_{I} - \hat{N}_{M,2}}(1-r_{I})^{\hat{N}_{M,2}}.
\end{equation}

Since $\hat{N}_{M,1}$ and $\hat{N}_{M,2}$ are drawn independently, 
we can represent the distribution of all measured galaxies using the triangular sum
\begin{equation}
    \begin{split}
&P(\hat{N}_{M}\mid r_{M}, r_{I}, N_{I},N_{M}) 
\\
&=\sum_{\hat N_M=\hat{N}_{M,1} + \hat{N}_{M,2}}P(\hat{N}_{M,1}\mid r_{M}, N_{M}) P(\hat{N}_{M,2}\mid r_{I}, N_{I}).
\end{split}
\end{equation}
or equivalently
\begin{equation}
\begin{split}
&P(\hat{N}_{M}\mid r_{M}, r_{I}, N_{I},N_{M}) 
\\
&= \sum_{\hat{N}_{M,1} = 0}^{\hat N_{M}}P(\hat{N}_{M,1}\mid r_{M}, N_{M}) P(\hat{N}_{M} - \hat{N}_{M,1}\mid r_{I}, N_{I}). 
\end{split}
\end{equation}
Additionally, since we know the total number of galaxies $N_\mathrm{tot}$ and are interested in the true underlying number of mergers $N_M$, we can write the likelihood as a function of $N_M$, $r_M$, and $r_I$,
\begin{equation}
\begin{split}
    \mathcal L(N_M,r_M,r_I&\mid \hat N_M)
    \\
    =P(\hat N_M&\mid r_M,r_I,N_\mathrm{tot}-N_M,N_M)
    \end{split}
\end{equation}

One benefit to this formalism is that it easily generalizes to an arbitrary number of classifiers, each with their own measurements and accuracies. Assuming that each classifier is independent, the set of all observations is distributed as
\begin{equation}
\begin{split}
    P(\{\hat N_{M,i}\}&\mid \{r_{M,i}\},\{r_{I,i}\},N_\mathrm{tot}-N_M,N_M)
    \\
    &=
    \prod_i
    P(\hat N_{M,i}\mid r_{M,i},r_{I,i},N_\mathrm{tot}-N_M,N_M)
\end{split}
\end{equation}
and we can write the likelihood as
\begin{equation}
    \begin{split}
    \mathcal L(N_M,\{r_{M,i}\},\{r_{I,i}\}&\mid\{\hat N_{M,i}\})
    \\ 
    = P(\{\hat N_{M,i}\}&\mid \{r_{M,i}\},\{r_{I,i}\},N_\mathrm{tot}-N_M,N_M).
\end{split}
\end{equation}

In this statistical model, classifiers' accuracies are nuisance parameters that need to be marginalized over since the true merger fraction is the variable of interest. Using Bayes' theorem, we write the posterior distribution
\begin{equation}
    \label{eq:posterior}
    \begin{split}
    P(N_M,&\{r_{M,i}\},\{r_{I,i}\}\mid\{\hat N_{M,i}\}) 
    \\
    &\propto P(\{\hat N_{M,i}\}\mid \{r_{M,i}\},\{r_{I,i}\},N_\mathrm{tot}-N_M,N_M)
    \\
    &\times P(N_M)P(\{r_{M,i}\},\{r_{I,i}\})
    \end{split}
    \end{equation}
where
$P(N_M)$ and $P(\{r_{M,i}\},\{r_{I,i}\})$ are the prior distributions.
We sample this posterior distribution using the Markov Chain Monte Carlo sampler \texttt{emcee}.\footnote{\texttt{emcee} is an implementation of the \citet{goodmanweare} Affine Invariant  MCMC Ensemble sampler \footnote{\url{https://emcee.readthedocs.io/en/stable/}\citep{emcee}}}

\begin{figure*}
    \centering
    \includegraphics[width=0.7\textwidth]{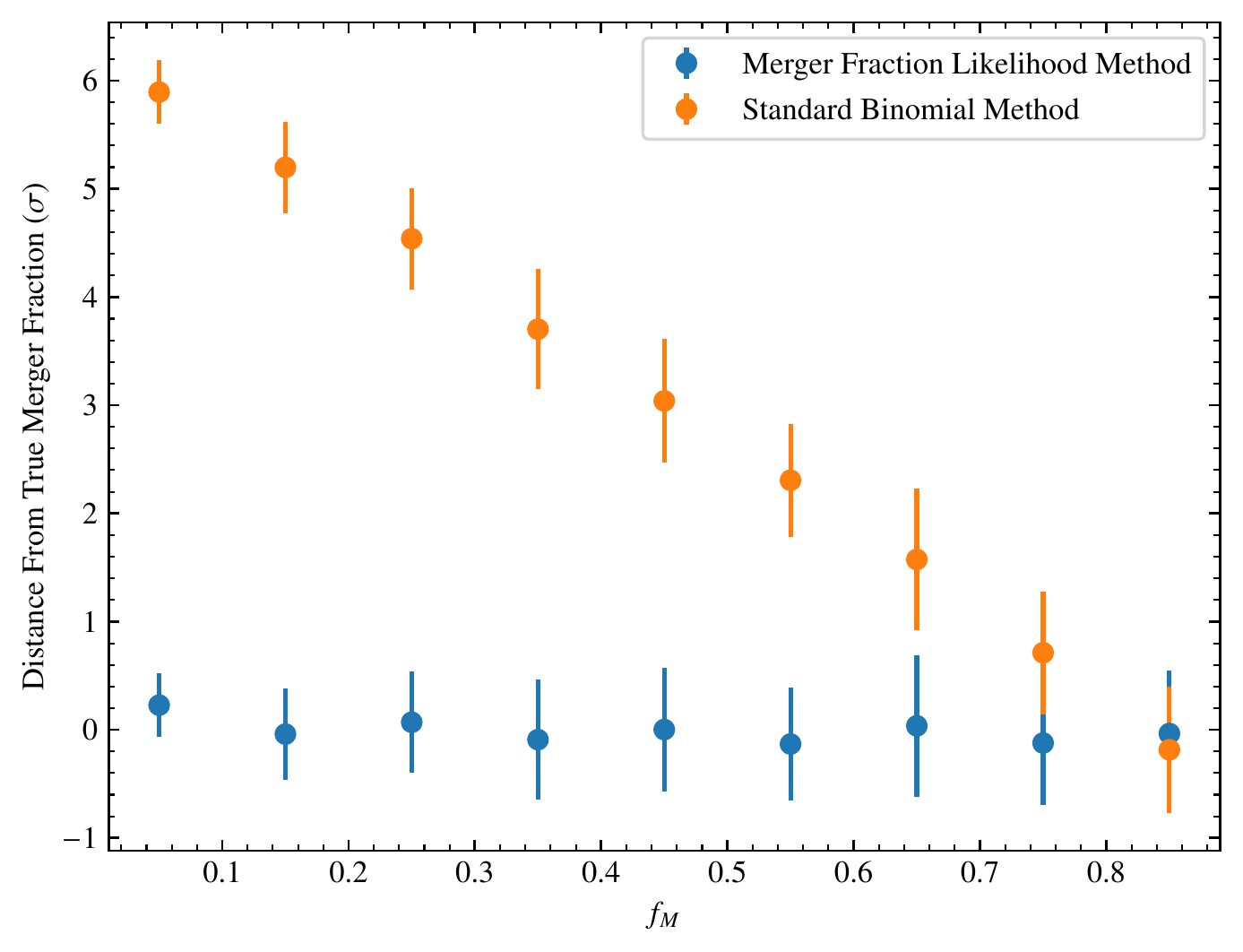}
    \caption{Recovery of Intrinsic Merger Fraction Using the Simulated Galaxy Catalogue Sample: Assuming the average accuracy of 14 simulated classifiers for merging and isolated systems is 80\% and 55\% respectively, we show how the sigma difference from the intrinsic fraction changes as a function of intrinsic merger fraction using the standard binomial method. The blue points are derived using our likelihood implementation. The orange points points use the standard binomial method. The error bars from each use the standard deviation from each method's own probability distribution function.}
    \label{fig:sim_binom_comp}
\end{figure*}

In principle, we can obtain the prior distribution of the classifiers' accuracies using their performances on mock galaxies whose underlying state is known a priori, although the applicability of this prior depends on (1) the number of mock galaxies and (2) the extent to which the mock galaxy sample can be treated as real data. 
We implement our prior using the classifiers' performance on the mock galaxies,
\begin{equation}
\begin{split}
    P(\{r_{M,i},r_{I,i}\})
    = \prod_i
    &P(r_{M,i}\mid \hat N_{M,\mathrm{syn},i}, N_{M,\mathrm{syn}})
    \\
    \times
    &P(r_{I,i}\mid \hat N_{I,\mathrm{syn},i}, N_{I,\mathrm{syn}})
\end{split}
\end{equation}
such that the classifier's accuracies are beta distributed such that $r_{M/I}\sim\mathrm{Beta}(\hat N_{M/I}+1,N_{M/I}+1)$.
In principle, we can apply an additional prior on $N_M$, $\{r_{M,i}\}$, and $\{r_{I,i}\}$, but we find that the full likelihood results are not noticeably affected by altering the prior.

The strength of this method is its internal consistency; given a set of observed mergers, $\{\hat  N_{M,i}\}$, the likelihood is maximized when a value of $N_M$ shown to all classifiers is most plausible, given a set of accuracies $\{r_{M,i},r_{M,i}\}$. This is in contrast to the usual approach, which assumes each classifier has perfect accuracy, and can only reflect reality if each classifier was shown a different set of galaxies. In \autoref{fig:stat_schema}, we show the graphical model of our likelihood analysis where all the variables are defined within this section.

\subsection{Testing the Likelihood Model on a Simulated Galaxy Catalogue} \label{sec:sim_gal}

To validate this model, we first simulate a data galaxy catalog with observations, best-fit values, uncertainties, and offsets from the input value;
\begin{itemize}
    \item Choose a true underlying merger fraction $f_M$, with $N_\mathrm{tot}$ galaxies, $f_MN_\mathrm{tot}$ mergers, and $(1-f_M)N_\mathrm{tot}$ isolated galaxies.
    \item Assign $n$ accuracy pairs $(r_{M,i},r_{I,i})$ drawn from a uniform distribution $\mathcal U(0.5,0.9)$ for each classifier, and calculate the mean accuracy for merging and isolated systems. Note the exact choice of the mean accuracies is unimportant for this exercise, but rather whatever choice is made is accounted for in the statistical modelling. 
    \item For each classifier, draw $\hat N_{M1,i}$ correctly identified mergers and $\hat N_{M2,i}$ incorrectly identified mergers, using the accuracies from the previous step.
\end{itemize}

In the standard binomial distribution approach, $nN_\mathrm{tot}$ galaxies have been observed, $\hat N_M=\sum_i \hat N_{M,i}$ mergers have been observed, and it is assumed that this observation is drawn from a binomial distribution,
\[
p(\hat N_M\mid nN_\mathrm{tot},f_M)=\binom{nN_\mathrm{tot}}{\hat N_M}
f_M^{\hat N_M}(1-f_M)^{nN_\mathrm{tot}-\hat N_M}.
\]
The likelihood $p(f_M\mid \hat N_M,nN_\mathrm{tot})$ is a beta distribution with parameters $\alpha=\hat N_M+1$ and $\beta=nN_\mathrm{tot}-\hat N_M+1$, and has mean and variance
\[
\frac{\hat N_M+1}{nN_\mathrm{tot}+2},\qquad
\frac{(\hat N_M+1)(nN_\mathrm{tot}-\hat N_M+1)}
{(nN_\mathrm{tot}+2)^2(nN_\mathrm{tot}+3)}.
\]

\begin{figure*}
    \centering
    \gridline{\fig{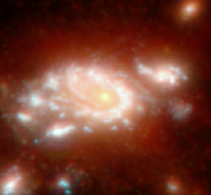}{0.4\textwidth}{(a) Noiseless Mock Galaxy} 
              \fig{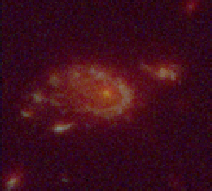}{0.4\textwidth}{(b) 3DHST GOODS-S \textit{HST} Survey \\ Noise Applied Mock Galaxy}}
    \caption{VELA+SUNRISE Noise-Added Mock Galaxy Example: In the left most image, we show the rgb (r: \textit{HST} WFC3/IR F160W, g: \textit{HST} ACS F775W, b: ACS F435W) VELA+SUNRISE image of a galaxy at redshift 1.7. This image has already been convolved with the \textit{HST} PSF in each of the wavelength bands. In the right most image, we show the same mock galaxy but with our noise model applied. The physical scale of the cutout is 7.8'' $\times$ 7.8'' or at z = 1.7, 67.2 kpc $\times$ 67.2 kpc  In the appendix we provide all merging and isolated mock galaxy noise applied images.}
    \label{fig:mock_eg}
\end{figure*}

In \autoref{fig:sim_binom_comp}, we compare the standard binomial approach against a test of the merger fraction likelihood model. We assume the average accuracies of 14 simulated classifiers for merging and isolated systems are 80\% and 55\% respectively. We then simulate 9 different data sets of 50 galaxies with an intrinsic merger fraction that spans from 0.05 to 0.95. We compare the distance in units of sigma from the true merger fraction for these different intrinsic merger fractions. The orange points represent the sigma difference from the truth for the standard binomial approach, and the blue points the sigma difference from the truth for the merger fraction likelihood method presented in this work. The difference between the standard binomial method from the intrinsic merger fraction varies as a function of the intrinsic merger fraction. In the case where the intrinsic merger fraction is 0.85, the standard binomial method is able to recover the true fraction due to the particular classifier accuracy values chosen in this test. In real classification studies, where the intrinsic merger fraction is a priori unknown, it is impossible to predict the deviation from the truth without accounting for the accuracy of the classifiers. The merger fraction likelihood method presented in this work takes into account the accuracy of the human classifiers.  As is seen in \autoref{fig:sim_binom_comp}, any biases inherent in our method should not significantly depend on the intrinsic merger fraction of the sample.

\subsection{Testing the Likelihood Model on Mock Galaxies with Real Human Classifiers} \label{sec:vela}


In this section, we detail a method where one can systematically estimate a real classifier’s accuracy using mock observations from a suite of galaxy formation simulations with known true classifications.




To do so, we use mock images created from the VELA zoom-in hydrodynamical galaxy formation simulations \citep{ceverino14,snyder15,simons19}. The VELA simulation suite comprises 35 galaxy halos, spanning virial masses of $\sim 1$--$20\times10^{11} M_{\odot}$ at $z = 2$. These simulations were run in a full cosmological context using the Adaptive Mesh Refinement Tree code ({\tt{ART}}; \citealt{kravtsov97}) and the subgrid physical recipes used are described in detail in \citet{ceverino10, ceverino12, ceverino14}.

For each timestep of each simulated VELA halo, the true classification (isolated or merging) of the central mock galaxy is determined from the kinematics and spatial distribution of its stars (described in \citealt{simons19}). Galaxies are selected as mergers if they have undergone a merger within the last 100 Myr, or if they have a companion galaxy within 35 kpc. We randomly select a set of 24 simulation outputs where the central galaxy is merging, and 29 simulation outputs where the central galaxy is isolated. These simulation outputs span redshifts of 1.0 to 3.5, and the redshift distribution of the isolated and merging galaxies are similar \autoref{tab:df_mock}. 

Mock {\emph{Hubble}} ACS/WFC3 images were created for each galaxy in the VELA suite in \citet{snyder15} and \citet{simons19}, using the dust-radiative transfer code {\tt{SUNRISE}} \citep{jonsson10}. The production of the mock images are described in detail in \citet{simons19}. The mock images are available as high level science products on a public repository.\footnote{\url{https://archive.stsci.edu/prepds/vela/}} 

We downloaded noise-free versions of the mock images of our selected mock galaxies in three {\emph{Hubble}} bands: ACS F435W, ACS F775W, and WFC3 F160W. The mock images include the appropriate spatial resolution and pixel scale of each band, but do not include noise. In addition to the 24 merging galaxies and 17 isolated galaxies, we create 10 images for a set of ''fake'' mergers. This set of ''fake'' mergers is used to assess how well a classifier can distinguish galaxies that are interpolating by chance alignment (i.e., not interacting) from galaxies that are merging. To do this, we superimpose the images of two mock isolated galaxies using a random separation less than 8$\arcsec$. 

We then add the appropriate amount of Poisson noise to simulate the well-studied real data-set of the 3DHST reduction of GOODS-South. We first calculate a normalization factor to match a background pixel in the VELA mock galaxy cutouts to the background in each \textit{HST} band of the 3DHST GOODS-S maps. We calculate the normalization factor by first performing aperture photometry on a real galaxy where the background is sky dominated. We multiply the aperture flux of the image by the exposure time, and get the instrument counts of the image. We then get the background counts from the \textit{HST} Exposure Time Calculator \footnote{http://etc.stsci.edu/etc/input/wfc3ir/imaging/}. 

\begin{figure*}
    \centering
    \includegraphics[width=1.5\columnwidth]{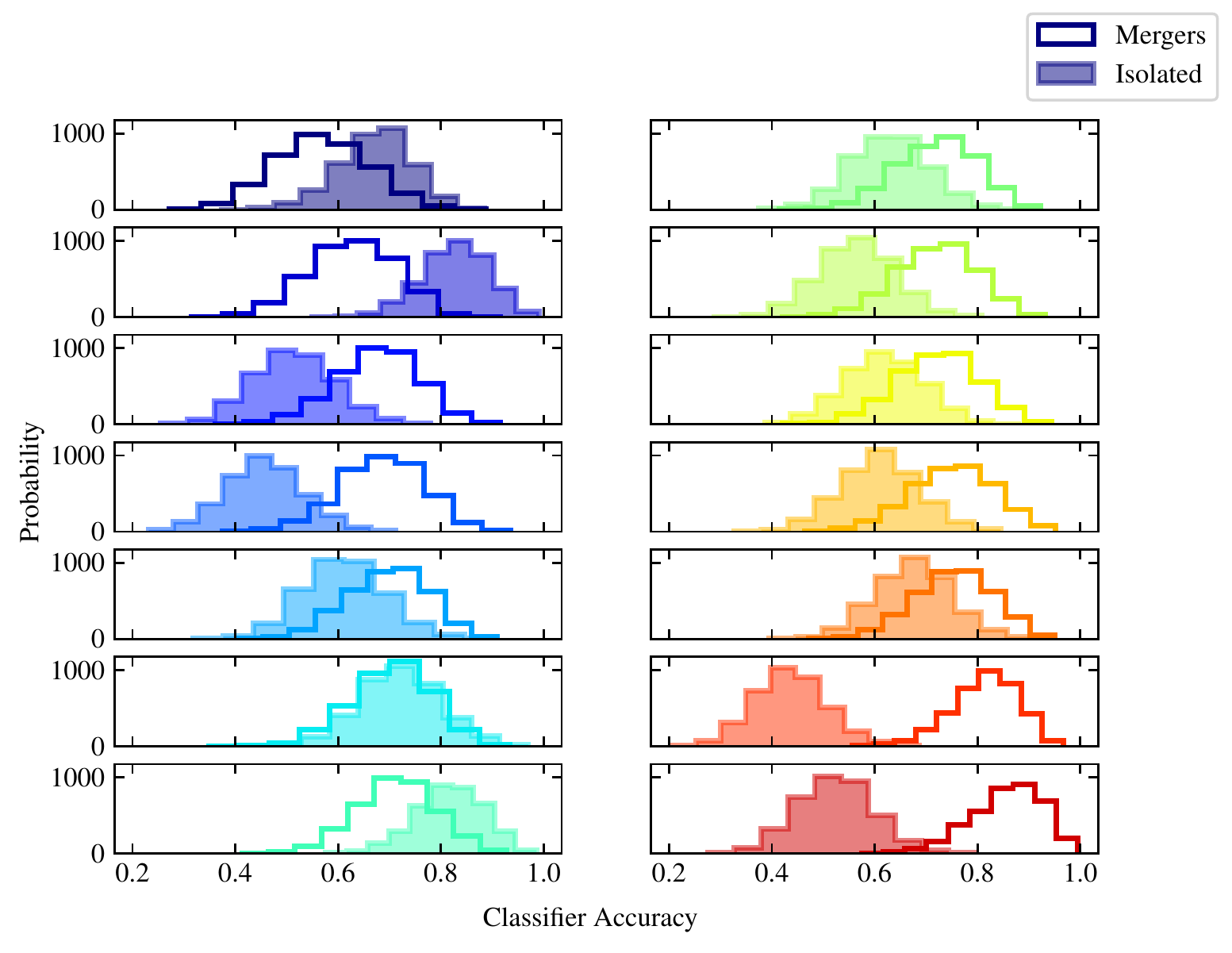}
    \caption{Estimation of the Merger and Isolated Classification Accuracies of Each Individual Human Classifier using their Classifications of the VELA+SUNRISE Noise-Added Mock Galaxy Sample: For each fourteen classifiers, we show the likelihood distribution for assessing isolated systems (filled histogram), and merging systems (unfilled histogram) using \autoref{eq:posterior}.}
    \label{fig:classigier_accuracy}
\end{figure*}

Next, we matched a VELA image to the real galaxy in redshift and flux for each individual \textit{HST} band. We again perform aperture photometry on the VELA image in all three bands. We finally apply the normalization factor of each band by multiplying the VELA image with the real image instrument counts divided by the VELA image counts. After pixel matching the VELA pixels to the 3DHST survey pixels we then apply additional sources of noise. Using IRAF's \texttt{mknoise} function we apply Gaussian read noise, gain, shot noise, and the background counts found from the ETC. In \autoref{fig:mock_eg} we show an example of a mock galaxy with and without the applied survey derived noise. In the left most image, we show the rgb (r: \textit{HST} WFC3/IR F160W, g: \textit{HST} ACS F775W, b: ACS F435W) VELA+SUNRISE image of a galaxy at z = 1.7. This image has already been convolved with the \textit{HST} PSF in each of the wavelength bands. In the right most image, we show the same mock galaxy but with our noise model applied. The physical scale of the cutout is 7.8'' $\times$ 7.8'' or at z = 1.7, 67.2 kpc $\times$ 67.2 kpc  In the appendix we provide all merging and isolated mock galaxy noise applied images. 

After creating the noise-added mock galaxy sample, we then showed fourteen different human classifiers the entire sample of mock images. The samples are intermixed, and are classified using the criteria enumerated below. The classifiers were also told there may be background or foreground galaxies in the images. The classifiers' backgrounds ranged from eight professors of astronomy, a post-doctoral fellow in astronomy, and four graduate students. The first author of this study was not included as a classifier as to minimize potential bias. We created a website where the mock images asked hosted, and asked each classifier to classify the image over the following options: 
\begin{enumerate}
    \item Merging: Major (approximately similar size) 
    \item Merging: Minor (approximately 1:4 size ratio) 
    \item Disturbance: Major 
    \item Disturbance: Minor 
    \item No Evidence of Merger/Interaction. 
\end{enumerate}

We provided the classifiers with the redshift of the central galaxy, and defined merging as an on-going interaction (which can include evidence of gravitational disturbances i.e., tidal tails with distinct galaxy systems, pairs). We defined a disturbance as a post-merger in the final stages of (or post-) coalescence. A disturbance classification can include large asymmetry/gravitational disturbance and/or tidal tails. Ultimately, for our analysis we use only two morphological classes: merging and not merging. Merging includes major mergers, minor mergers, and major disturbances. The non-merging class includes minor disturbances and no-evidence of gravitational interactions. This is due to the difficulty in constraining merger stage and mass ratio from images alone. Nonetheless, when the human classifiers are presented with the images they are given multiple morphological divisions to choose from to help aid in the human classification process.

We then use the raw accuracies of the classifiers to inform a data driven model of determining the merger fraction of the sample. In the simulated galaxy case, we assume perfect knowledge of the accuracies of each classifier, or a $\delta$-function prior for each accuracy parameter that is the same as the input value. For the real human classifications on the VELA+SUNRISE noise-added mock galaxy sample, we estimate the accuracies from the mock images, where $r_M=\hat N_{M,s}/N_{M,s}$ and $r_I=\hat N_{I,s}/N_{I,s}$. We collapsed the classification options of the mock galaxies in two options: merging and non-merging. Merging includes major mergers, minor mergers, and major disturbances. The non-merging class includes minor disturbances and no-evidence of gravitational interactions.  

\begin{figure}
    \centering
    \includegraphics[width=.5\textwidth]{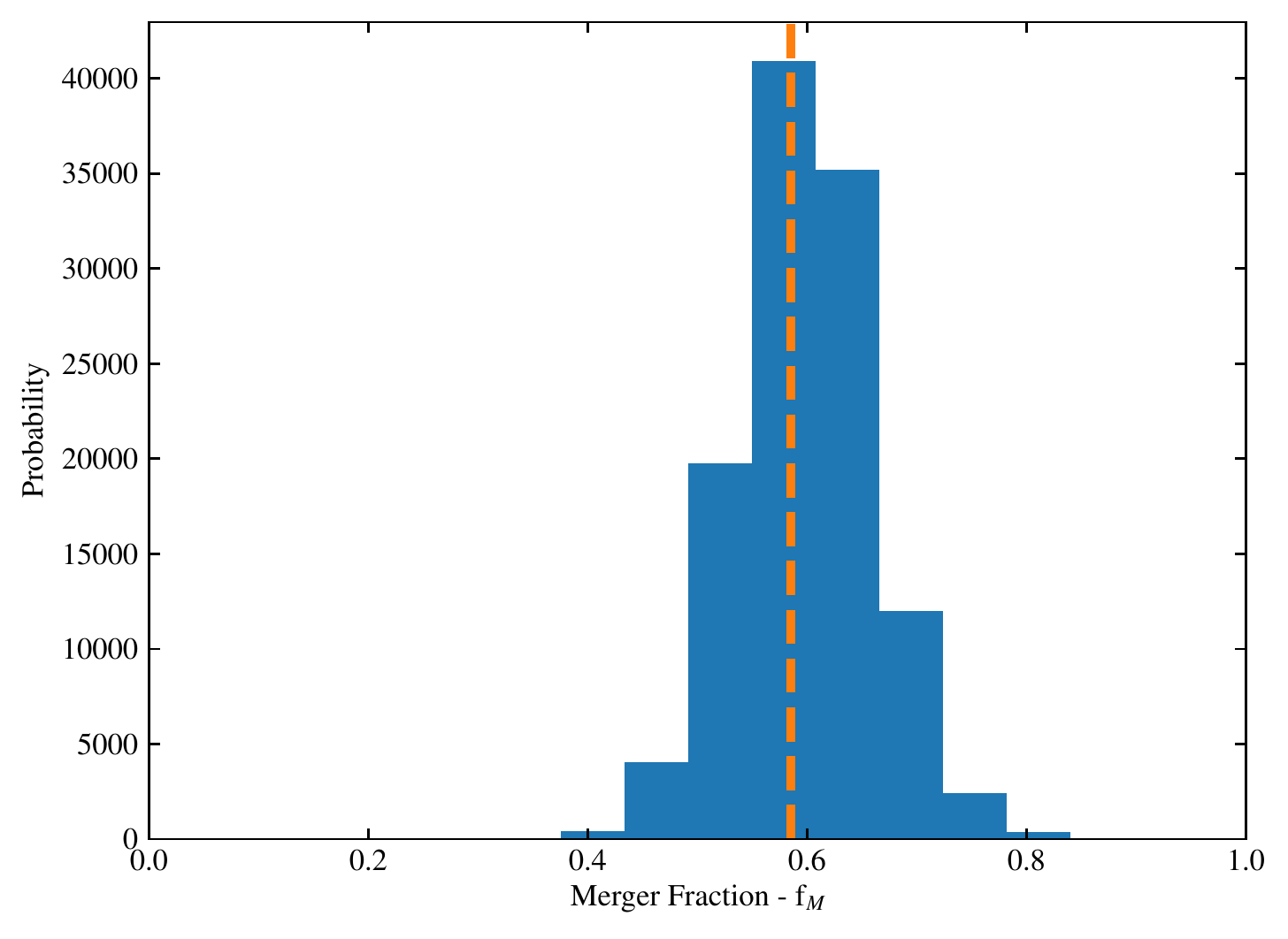}
    \caption{Measured Merger Fraction of the VELA+SUNRISE Noise-Added Mock Galaxy Derived from the Likelihood Model: The dashed, orange line is the intrinsic merger fraction of the mock galaxy sample. The blue histogram is the probability distribution derived from the likelihood model or equation 11, with mean 0.58 $\pm$ 0.06.}
    \label{fig:mock_merger_frac_likelihood}
\end{figure}

In \autoref{fig:classigier_accuracy}, we show the estimation of the merger and isolated classification accuracies for each individual classifier. As shown in \autoref{eq:posterior}, the classifier accuracies are estimated using the raw accuracies from the mock galaxy classifications and the individual agreement on the number of galaxies in a merger in the mock galaxy sample. We show the likelihood distribution for assessing isolated systems (filled histogram) and merging systems (unfilled histogram). We find that some classifiers have higher accuracies assessing isolated systems, some have higher accuracies assessing merging systems, and some that are equally accurate for both. As mentioned in \autoref{sec:problem}, the effect of a classifiers bias for or against a specific morphological class depends on the intrinsic merger fraction of the population. The fourteen classifiers chosen have a diverse range of accuracies, and a standard binomial statistical approach would not capture this significant source of error. Using the likelihood model, we recover a total merger fraction of 56.8\%$\pm 0.06$ (25/41) at the 95\% confidence level. We are well within 1$\sigma$ of the true merger fraction of the mock sample which is 54.5\% (24/41). In \autoref{fig:mock_merger_frac_likelihood}, we show the probability distribution of the merger fraction for the mock galaxy sample. The dashed, orange line is the intrinsic merger fraction of the mock galaxy sample. The blue histogram is the probability distribution derived from the likelihood model or equation 11, with mean 0.57 $\pm$ 0.06.

There are some important caveats with this approach and implementation. First, we do not know if classifiers will characterize the mock galaxies the same way as they do real images. Though, our algorithm is developed such that any metric of estimating a classifier's accuracy can be used instead. Second, we use a point estimate, the raw merger fraction of each classifier, when it would be more appropriate to use a beta distribution prior in our fits. We test whether these affects will significantly bias our results, and  we find when we run the analysis on the mock galaxies, we recover the input merger fractions correctly, and the fit has a similar likelihood surface. We also find when we run a full Monte Carlo Markov Chain with flat priors on the accuracies, the output mean and variance are consistent with the mock image estimate within a standard deviation. 

Most importantly, a drawback to this method is the comparison of aggregates rather than individual galaxies. For example, two classifiers could disagree on which specific galaxies are in mergers, but find similar merger fractions in the sample. The aggregate method could erroneously imply that the two classifiers agree on their classifications, when in fact they do not.

\section{A New Approach: Simultaneously Estimating the Merger Fraction and the Probability of an Individual Galaxy Being in a Merger} \label{sec:per_gal}

Another way to statistically estimate the merger fraction of a sample is to include the level of agreement, or the amount of classification agreement between individual classifiers on a given galaxy, to estimate the merger fraction probability of a sample. In this section, we construct a method that  accounts for the accuracy of a classifier using the level of individual galaxy agreement in addition to their assessment on mock images. In this new method, we are able to simultaneously estimate the merger fraction of a population and the probability of each individual galaxy being in a merger.  

\subsection{Statistical Framework for Per Galaxy Merger Assessments}

Only a few recent works have had enough human classifiers to assume a frequentist approach and use the mean of the individual classifications of a given galaxy to estimate the individual galaxy morphology (i.e., \textit{GalaxyZoo} 1 and 2, \citealt{galaxyzoo1,galaxyzoo2}). In \textit{Galaxy Zoo} 2, their main sample of 283,971 galaxies had a median of 44 classifications; the minimum was 16, and $>99.9\%$ of the sample had at least 28 classifications. Even in the case of many individual classifications of a given galaxy, it is unclear what minimum number of classifications is needed in order to ignore intrinsic merger fraction dependent biases. 

Since a large fraction of merger studies have smaller samples and consequently less human classifiers, often there are not enough individual classifications on a given galaxy to robustly report a classification and error of the classification for that galaxy in the manner that \textit{Galaxy Zoo} studies can. In this new approach, we constrain an individual's accuracy (similar to the method presented in \autoref{sec:merger_frac_like}), and using this information we show we can estimate the merger fraction of a sample and the probability of an individual galaxy being in a merger. 

If a respondent is shown a merger, they will say it is a merger with probability $r_M$, or say it is isolated with probability $1-r_M$. Conversely, if it is isolated, they will say it is a merger with probability $1-r_I$ or say it is isolated with probability $r_I$. Thus respondent $i$ classifies $j$th galaxy $G$ with classification $m$ as
\begin{equation}
    p(m_i\mid G_j)
    =
\begin{cases}
    r_M &   m_i=G_j=\mathrm{merger}\\
    1-r_M&m_i\neq G_j=\mathrm{merger}\\
    r_I&m_i=G_j=\mathrm{isolated}
    \\
    1-r_I&m_i\neq G_j=\mathrm{isolated}
    \end{cases}
\end{equation}

With more sub-categories, this can be generalized to $p(m_i\mid G_j)=r_{ij}$, where $\sum_i r_{ij}=1$. 
The mock galaxy sample presented in \autoref{sec:merger_frac_like} is a Bernoulli trial, although technically the respondents were asked to choose one option out of five. The generalization is described by a multinomial distribution, and its conjugate distribution is Dirichlet.

The likelihood of the classifications of a single galaxy by multiple classifiers given a merger fraction and classifier accuracies can be written
\begin{equation}
\begin{split}
    p(\{m_i \} \mid \{r_i\}, f_m) = f_m \prod_{i} p(m_i\mid G={\mathrm M}) \\ 
    + (1-f_m) \prod_{i} p(m_i\mid G={\mathrm I}).
\end{split}
\end{equation}

In this expression, the true nature of the galaxy in question is marginalized out.
Expanding to multiple galaxies, we get the likelihood for the classifications of a collection of galaxies:
\begin{equation}\label{eq:pergal_like}
    p(\{m_{ij}\} \mid \{r_i\}, f_m) = \prod_{j} p(\{m_{ij}\} \mid \{r_i\}, f_m).
\end{equation}

Multiplying this likelihood by a prior on the merger fraction and, if the classifier accuracies are not held fixed, by a prior on accuracies gives the unnormalized posterior probability distribution function for this model.

If we wish to recover the probability that a particular galaxy is a merger, we can use the expression
\begin{equation} \label{eq:per_gal}
    p(G = M \mid \{m_i\}, \{r_i\}, f_m) = \frac{f_m \prod_i p(m_i \mid G=M)}
    {p(\{m_i\} \mid \{r_i\}, f_m)}.
\end{equation}
The probability that this galaxy is isolated is the complement of this expression. The classifier's observations of simulated galaxies can be used as a prior on the observer's accuracies $r_{M/I}$, depending on how they classify the known synthetic population. This gives an informative prior, which inherently assumes that the synthetic catalog is statistically similar to the real catalog.

The strength of this method is its internal consistency; given a set of observed mergers, the likelihood is maximized when a value of $f_M$ shown to all classifiers is most plausible given a set of individual classifications for each galaxy. We evaluate \autoref{eq:pergal_like} using the Markov chain Monte Carlo No-U-Turn Sampler algorithm \citep[details within][]{Hoffman2014} using the open source probabilistic programming framework \texttt{PyMC3} \citep[][]{pymc3}. 

The likelihood function of a given galaxy having a specific morphological classification requires a robust statistical description of a human classifiers accuracy in assessing both merging and isolated systems. In the previous step, where we maximize the likelihood of a population's merger fraction, our algorithm also maximizes the likelihood of an individual galaxy's classification. This allows for deeper data exploration on galaxy samples that are normally too small to do anything but population averages. 

We also provide to the community the full code repository to calculate the merger fraction probability and probability of an individual galaxy being in a merger given a set of individual galaxy classifications and an estimate of the classifier accuracy. \footnote{https://github.com/elambrid/merger\_or\_not}

\begin{figure}
    \centering
    \includegraphics[]{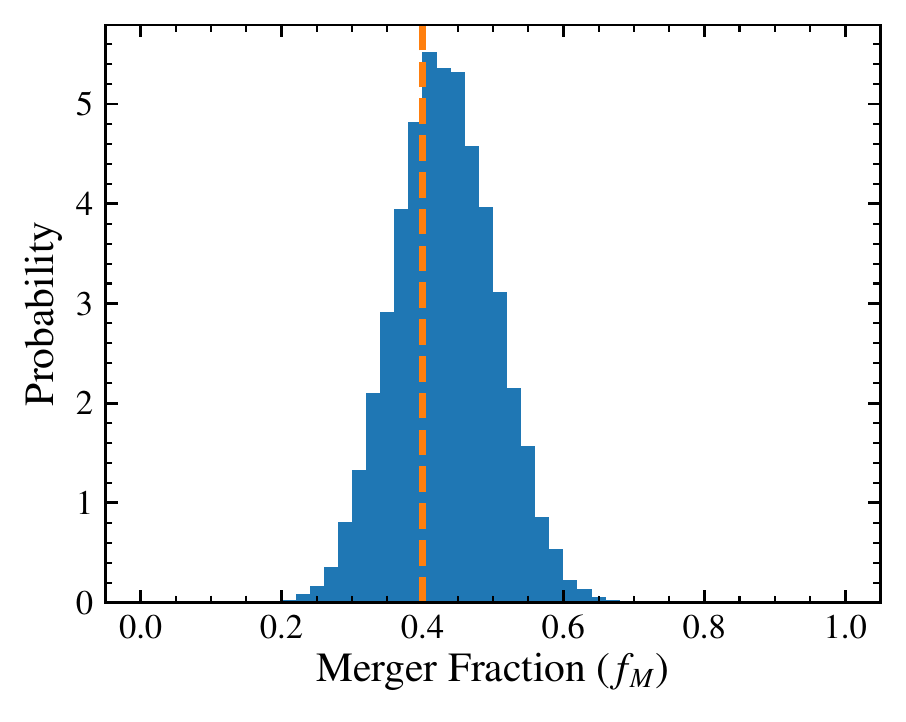}
    \caption{Merger Fraction Probability for a Simulated Galaxy Catalogue Using Level of Classifier Agreement per Galaxy: The orange dashed lines corresponds to the merger fraction truth value of 0.4. The blue histogram is the merger probability of a simulated galaxy catalogue (50 objects) using simulated classifications (14 classifiers) with mean accuracies r$_M$=0.75 and r$_I$=0.65 for identifying mergers and isolated galaxies respectively.
    }
    \label{fig:simulated_pergal_frac}
\end{figure}

\begin{figure}
    \centering
    \includegraphics[]{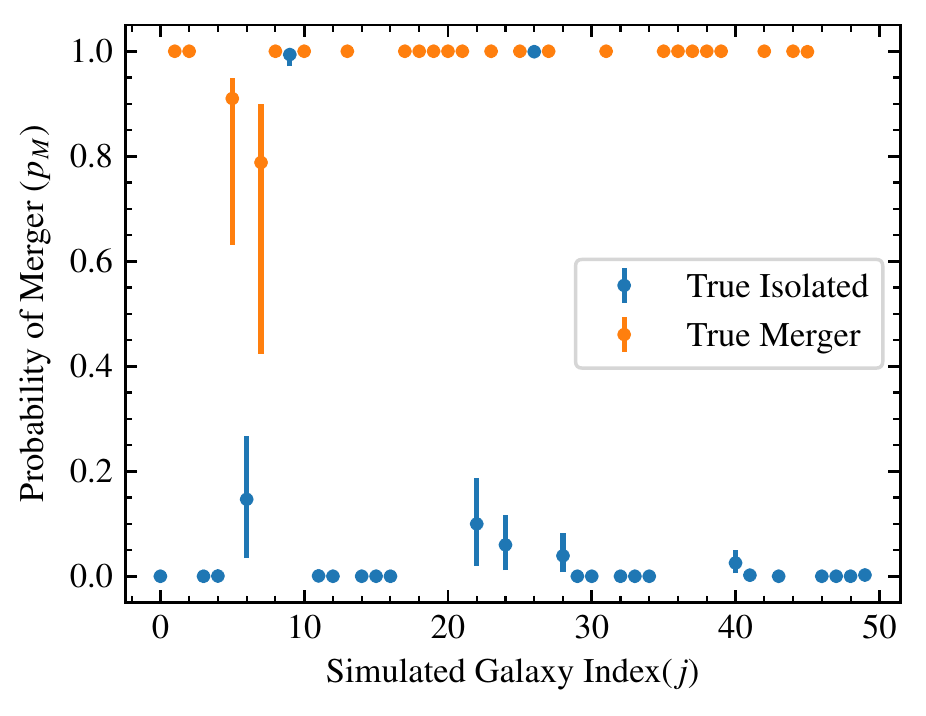}
    \caption{Probability of Individual Galaxy Being in a Merger for a Simulated Galaxy Catalogue: The simulated galaxies are labeled by simulated classifiers (14) with the same perfectly known $r_M$ and $r_I$ values used in \autoref{fig:simulated_pergal_frac}.
    }
    \label{fig:simulated_pergal_pergal}
\end{figure}

\subsection{Testing Per Galaxy Model on Simulated Data}

Similarly as in \autoref{sec:sim_gal}, we simulate a galaxy catalog with imaginary classifications for each individual galaxy to obtain the probability that each galaxy is in a merger. We test with a true underlying merger fraction $f_M=0.4$. We randomly assign 14 accuracy pairs from a uniform distribution $r_M\sim \mathcal U(0.5,0.9)$ and $r_I\sim\mathcal U(0.5, 0.9)$ for each classifier. For each classifier, we assign 50 observations, and use mean accuracies r$_M$=0.75 and r$_I$=0.65 for identifying mergers and isolated galaxies respectively. In \autoref{fig:simulated_pergal_frac}, we show the merger fraction probability for the simulated galaxy catalogue using the statistical framework presented in \autoref{sec:per_gal}.  The mean estimated merger fraction probability is within a $\sigma$ of the true merger fraction of 0.4: $f_M=0.43 \pm 0.07$. As a consistency test, we evaluated the likelihood function with 50 different randomly generated galaxy catalogues and classifier accuracies, and we find, for every test, the mean merger fraction is within 1$\sigma$ of the input true merger fraction. 

As shown in \autoref{eq:per_gal}, we can also estimate the probability of an individual galaxy being in a merger. Using the same example parameters as in \autoref{fig:simulated_pergal_frac}, we show the probability of 50 galaxies being in a merger given the above simulated set-up in \autoref{fig:simulated_pergal_pergal}. We find two galaxies that are mis-classified, which yields an overall accuracy of 96\%.



\subsection{Testing on Mock Galaxies with Real Human Classifiers}

\begin{figure*}
    \centering
    \includegraphics[scale=0.8]{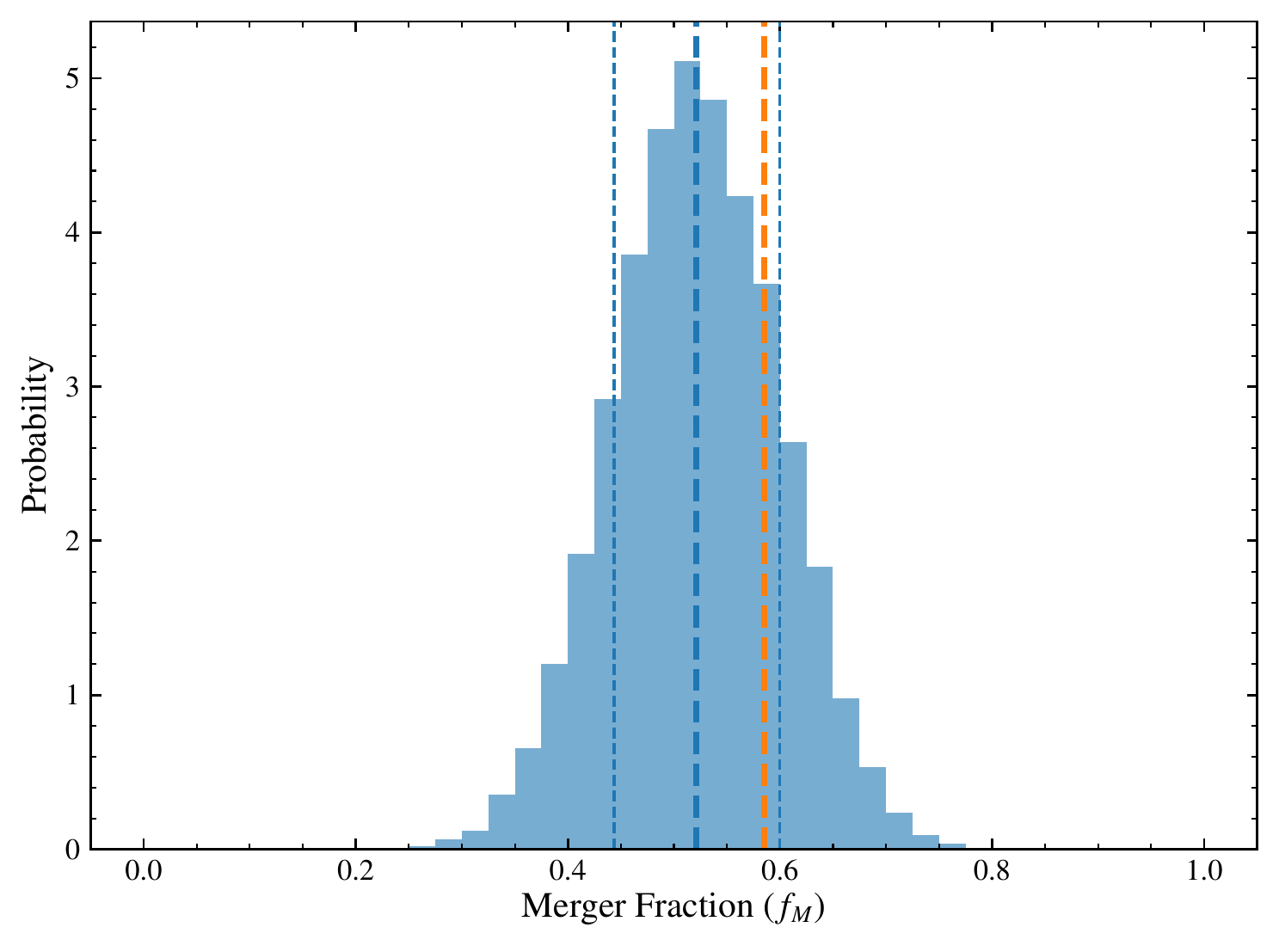}
    \caption{Merger Fraction Probability for a sample of Mock Galaxies Classified by Real Humans: The orange dashed line corresponds to the merger fraction truth value of 0.59. The blue histogram is the merger probability of the mock galaxy catalogue (41 objects) using real human classifications (14 classifiers) with the accuracies at identifying mergers and isolated galaxies estimated using \autoref{eq:pergal_like}.}
    \label{fig:mock_pergal_frac}
\end{figure*}

\begin{figure}
    \centering
    \includegraphics[]{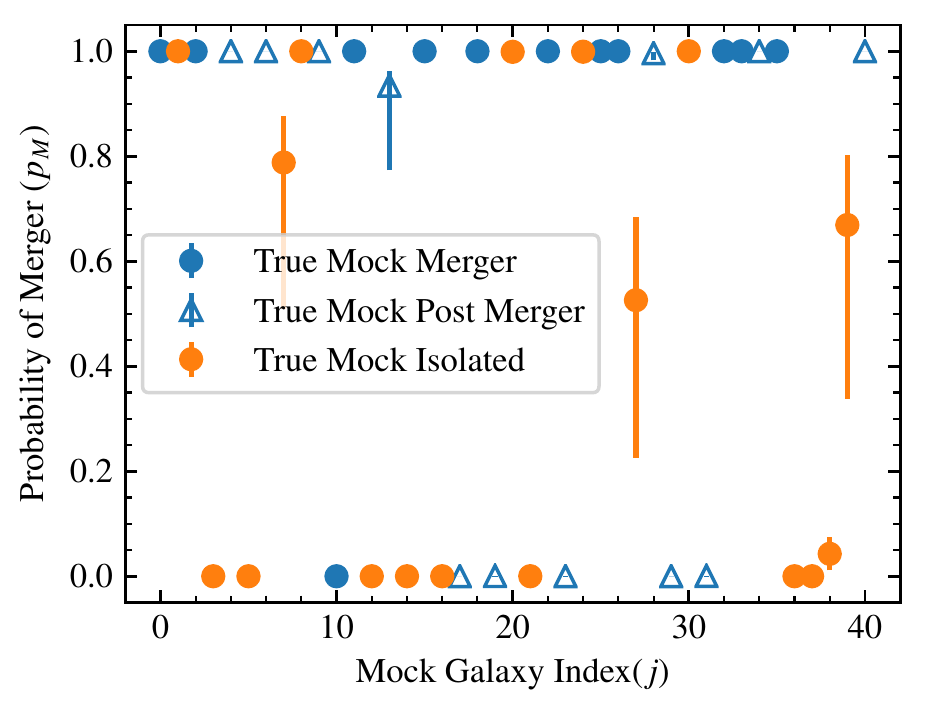}
    \caption{Probability of Individual Mock Galaxies Being in a Merger: The mock galaxies are classified by  14 real human classifiers with a range of accuracies estimated using \autoref{eq:pergal_like}. The solid blue circles correspond to the mock galaxies undergoing a merger, the empty blue triangles are mock galaxies that have coalesced within 100 Myrs, and the orange filled circles are mock galaxies in isolated systems.)
    }
    \label{fig:mock_pergal_pergal}
\end{figure}

We now test how well our method recovers the properties of mock galaxies as observed by real classifiers. This is an important test, as the mock images are constructed to be realistic, and the ability of classifiers to identify them correctly should be closely related to classifiers' ability to identify real galaxies' properties. In addition, this test allows us to look at the failures of the model on a per-image basis and determine whether the issue comes from the algorithm or the data.

Using the formalism of this section, we can estimate the probability that each mock galaxy is a merger given the accuracies of our classifiers and their agreement on classifications of individual galaxies. We use the full Monte Carlo Markov Chain samples to estimate the scatter in this value. In other words, for each step in the chain \texttt{i}, there is a vector of parameters $N_{M,\mathtt{i}}$, $\{r_{M,i,\mathtt{i}}\}$, and $\{r_{I,i,\mathtt{i}}\}$ that can be used to estimate the probability of a galaxy being in a merger for that step, $p_\mathtt{i}(\mathrm{merger}\mid \{m_{ij},r_{M,ij,\mathtt{i}}, r_{I,ij,\mathtt{i}}\})$. We can then report the probability of this galaxy being in a merger by finding the 5\%, 50\%, and 95\% percentiles, corresponding to $2\sigma$ Gaussian errors.

Using the above methodology, we calculate the per-galaxy merger probability for the mock galaxy sample. We simultaneously estimate the probability of the merger fraction, the accuracies of the classifiers, and the probability that each mock galaxy is in a merger given every classifier's label. We first separate the ''fake mergers'' from the merging and isolated mock galaxy sample. We do this to strictly test the ability of human classifiers to identify objects that are not intrinsically merging but have small on-sky separations. We found 50\% of the ''fake mergers'' were erroneously classified as mergers, and 100\% of sources that were within 2'' of each other were erroneously classified as mergers. Thus, even with the usage of mock galaxy samples aimed to mimic the color differences due to objects at different redshifts, in this test, humans struggled to interpret their merger status correctly. We do not include the ''fake mergers'' in the merger fraction analysis presented in the rest of this section. To truly construct a test in the context of the effect of ''fake close pairs'' on the error of the merger fraction the entire mock galaxy sample would need to incorporate fake galaxies at various redshifts $>$8'' of the central source in such a fashion to simulate real galaxy cut-outs. While this is an important exploration of a potential source of human bias on merger fraction measurements, it is outside the scope of this study and will be incorporated in future work.  


In \autoref{fig:mock_pergal_frac}, we show the merger fraction probability for a sample of merging and isolated mock galaxies classified by real humans. The orange dashed line corresponds to the merger fraction truth value of 0.59. The blue histogram is the merger probability of the mock galaxy catalogue (41 objects) using real human classifications (14 classifiers) with the accuracies at identifying mergers and isolated galaxies estimated using \autoref{eq:pergal_like}. The mean estimated merger fraction probability is within a $\sigma$ of the true merger fraction of 0.59: $f_M=0.52 \pm 0.08$. In \autoref{fig:mock_pergal_pergal}, we show the probability of 41 galaxies being in a merger given the above simulated set-up. We find 6 galaxies that are mis-classified as mergers ($p_{M} \geq 0.8$, which yields a merger accuracy of 85\%.

We can understand the implications of \autoref{fig:mock_pergal_pergal} by determining the completeness, and overall accuracy of our model at inferring true merger classifications in the context of their merger state. We define completeness as the intrinsic merging mock galaxies with $p_{M} \geq 0.8$ compared to the total amount of intrinsic mock merging galaxies, and accuracy as the comparison between the estimated merger fraction and the true merger fraction of the mock galaxy sample. Using the classified mock galaxy intrinsically merging sample, we split the classifications of intrinsically pre-coalesced vs intrinsically post-coalesced (post-merger) systems. Note that, a mock galaxy is defined as a post-coalesced system if in the previous time-stamp, or 100 Myrs prior, the system was undergoing a merger with at least a mass ratio of 0.25.By our framework, pre-coalesced and post-coalesced systems are both defined as ''mergers'', and we aim to test whether intrinsically post-coalesced systems have similar classification accuracies as pre-coalesced systems.

We find 15\% of the intrinsically merging systems are mis-classified as isolated galaxies, thus a completeness of 85\%. We note all but one of the mis-classified mock galaxies are post-mergers. Upon deeper inspection, all of the falsely categorized mock post-mergers are in the top 50\% of the mock merger redshift distribution with mean and median $z = 1.98$, $z = 2.03$ respectively. When removing post-merging galaxies with $z > 2.0$, we find a completeness of 90\% and total accuracy of 92\%. Our results show that human visual classification can identify post-merging systems within 100 Myrs of coalescence up to $z=2.0$ with 92\% accuracy. For on-going merging galaxies, using our model, humans are able to robustly classify on-going significant mergers up to $z=3.0$ with 92\% accuracy. The decrease in accuracy of human classified post-merging systems is not surprising because merger features become more faint as time from coalescence increases. Previous studies that have combined pre- and post-coalesced merging galaxies across a large redshift range may be particularly susceptible to under-estimating the overall merger fraction of their sample. Due to the relatively course time resolution of VELA-Sunrise snapshots, we are unable to further test the effect of merger feature dimming post-coalescence.

We also note five isolated galaxies are erroneously measured as mergers. When removing higher red-shift post-merging galaxies from the sample, these mis-classified isolated systems drive the inaccuracy of our results. These five galaxies in particular have 95\% agreement of a ''merging'' classification from the 14 human classifiers. Future work will consist of understanding how the accuracy of human classification varies as a function of additional galaxy properties (i.e stellar mass, minor mergers) to understand why there can be such high classifier agreement on mis-classified sources.




\section{Summary and Conclusions}
\label{sec:conclusions}

In this work we propose a method of quantifying and accounting for merger biases of individual human classifiers and incorporate these biases into a full probabilistic model to determine the merger fraction of a population, and the probability of an individual galaxy being in a merger. 
We find the bias introduced from human classification is dependent on the intrinsic merger fraction of the population, and thus in order to report robust results from human visually classified data-sets, the bias from humans must be quantified. 

We then construct a likelihood model to determine the merger fraction of a sample given a set of human classifications. We apply this model using two different data-sets: (1) A simulated galaxy catalogue with simulated classifications (2) Real Human Classifications on a sample of mock galaxies derived from the VELA-SUNRISE sample, a catalogue of zoom-in hydro-dynamical galaxy simulations with synthetic Hubble ACS/WFC3 images \citet{simons19}. We recover the merger fractions to within 1\% of the truth for the simulated galaxy catalogue with simulated classifiers. For the real human classifications on a sample of mock galaxy images, we recover the merger fraction to within 1\% of the true merger fraction.


We then create a model to simultaneously determine the merger fraction, human accuracies and probability of each individual galaxy being in a merger. Using simulated human responses and accuracies, we are able to correctly label a galaxy as a ''merger'' or ''isolated'' to within 3\% of the truth. Using the mock galaxies with real human classifications, our model is able to recover the pre-coalescing merger fraction to within 10\%. For galaxies that have coalesced within 100 Myrs, our model recovers the intrinsic merger fraction to within 10\% for the sources that occupy the lowest 50\% of the redshift distribution. For the post-coalesced sources in the top 50\% of the redshift distribution (i.e $z \sim 2.0$), the accuracy of human classifiers significantly drops, and our model infers a merger fraction within 15\% of the truth. Note, this specific bound is observed at this redshift due to the mock galaxy images incorporating a noise model that will reflect the sensitivity of GOODS-S Hubble Observations. Thus, this important estimate on human classifier accuracy must be incorporated in merger studies that contain high redshift post-merger sources in the GOODS-S field.  

The implementation of our Bayesian model in studies that assess the merger state of $0.5 < z < 2$ galaxies using human classifiers yields better understood errors on the merger fraction. In addition, this statistical framework is able to more robustly constrain the probability of individual galaxies being in mergers with a smaller number of human classifiers than was previously possible.

\acknowledgments

We thank the anonymous referee for their thoughtful insight and important contributions to this work. In addition, we thank [insert] for useful discussions and insight. ELL is supported by [].

\software{\texttt{astropy} \citep{astropy:2013,astropy:2018},
          \texttt{corner} \citep{corner},
          \texttt{emcee} \citep{emcee},
          \texttt{IPython} \citep{ipython},
          \texttt{matplotlib} \citep{matplotlib},
          \texttt{numpy} \citep{numpy},
          \texttt{pandas} \citep{pandas, pandas2},
          \texttt{scipy} \citep{scipy},
          \texttt{statsmodels} \citep{statsmodels}
        }



\appendix

\section{Mock Galaxy Images and Classifications}

\begin{figure*}\label{fig:pre_mergers}
\begin{tabular}{cccc}
\subfigure[5: correct]{\includegraphics[width = 1.5in]{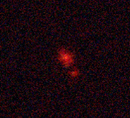}} &
\subfigure[7: correct]{\includegraphics[width = 1.5in]{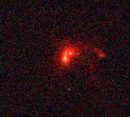}} &
\subfigure[12: incorrect]{\includegraphics[width = 1.5in]{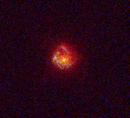}} &
\subfigure[37: correct]{\includegraphics[width = 1.5in]{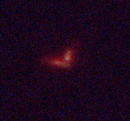}}\\
\subfigure[53: correct]{\includegraphics[width = 1.5in]{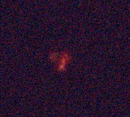}} &
\subfigure[38: correct]{\includegraphics[width = 1.5in]{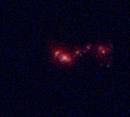}} &
\subfigure[16: correct]{\includegraphics[width = 1.5in]{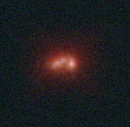}} &
\subfigure[41: correct]{\includegraphics[width = 1.5in]{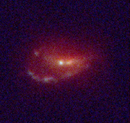}}\\
\subfigure[50: correct]{\includegraphics[width = 1.5in]{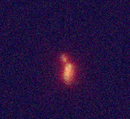}} &
\subfigure[36: correct]{\includegraphics[width = 1.5in]{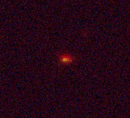}} &
\subfigure[25: incorrect]{\includegraphics[width = 1.5in]{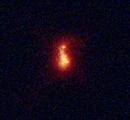}} &
\subfigure[28: correct]{\includegraphics[width = 1.5in]{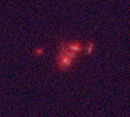}}
\end{tabular}
\caption{Pre-coalescence VELA+SUNRISE Noise Added Mock Merging Galaxies. The ID and measured classification are provided for each image (r:HSTWFC3/IR F160W, g:HSTACS F775W, b: ACS F435W). If a galaxy is classified correctly , p$_{M} > 99\%$, we identify it as ''correct''. All cutouts are 8''x 8''.}
\label{fig:pre}
\end{figure*}

\begin{figure*}\label{fig:post_mergers}
\begin{tabular}{cccc}
\subfigure[46: correct]{\includegraphics[width = 1.5in]{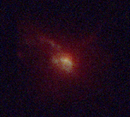}} &
\subfigure[49: correct]{\includegraphics[width = 1.5in]{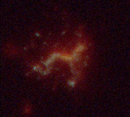}} &
\subfigure[40: correct]{\includegraphics[width = 1.5in]{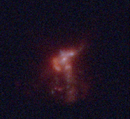}} &
\subfigure[32: correct]{\includegraphics[width = 1.5in]{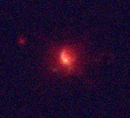}}\\
\subfigure[24: incorrect]{\includegraphics[width = 1.5in]{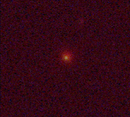}} &
\subfigure[44: incorrect]{\includegraphics[width = 1.5in]{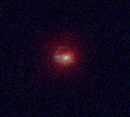}} &
\subfigure[31: incorrect]{\includegraphics[width = 1.5in]{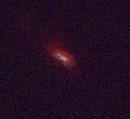}} &
\subfigure[15: correct]{\includegraphics[width = 1.5in]{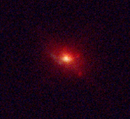}}\\
\subfigure[33: incorrect]{\includegraphics[width = 1.5in]{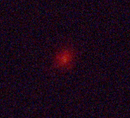}} &
\subfigure[8: incorrect]{\includegraphics[width = 1.5in]{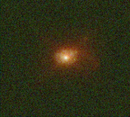}} &
\subfigure[6: correct]{\includegraphics[width = 1.5in]{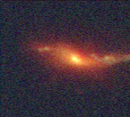}} &
\subfigure[11: correct]{\includegraphics[width = 1.5in]{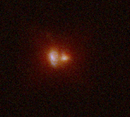}}\\
\end{tabular}
\caption{Post-coalescence VELA+SUNRISE Noise Added Mock Merging Galaxies. The ID and measured classification are provided for each image (r:HSTWFC3/IR F160W, g:HSTACS F775W, b: ACS F435W). If a galaxy is classified correctly, p$_{M} > 99\%$, we identify it as ''correct''. All cutouts are 8''x 8''.}
\label{fig:post}
\end{figure*}

As described in \autoref{sec:vela}, \autoref{fig:pre}, \autoref{fig:post}, \autoref{fig:iso} comprise the VELA+SUNRISE noise-added merging and isolated mock galaxy sample. The images are identified by their ID number. We also label each image as ''correct'' or ''incorrect''. Using \autoref{eq:per_gal}, we label a galaxy as ''correctly'' measured if the probability of being in a merger,p$_{M}$, is $>$ 99\% and the error on the probability is less than 10\%. In \autoref{tab:df_mock}, we list the z, intrinsic type, and measured probability of  each VELA+SUNRISE noise-added mock galaxy.  

\begin{table*}[ht]
\begin{center}
\subtable{
\begin{tabular}[t]{|c|c|c|c|c|c|}
\hline
ID & VELA ID& z & Type & p$_{M}$ & err \\ \hline
0  & 32,320,10 & 2.13 & i & 0.00 & n \\ \hline
1  & 7,370,10  & 1.70 & i & 1.00 & n \\ \hline
2  & 27,330,10 & 2.03 & i & 0.53 & y \\ \hline
3  & 20,290,5 & 2.45 & f & 1.00 & n \\ \hline
4  & 4,460,10  & 1.17 & i & 1.00 & n \\ \hline
5  & 2,290,10  & 2.45 & m & 1.00 & n \\ \hline
6  & 29,480,10 & 1.08 & m & 1.00 & n \\ \hline
7  & 6,310,10  & 2.23 & m & 1.00 & n \\ \hline
8  & 33,330,5  & 2.03 & m & 0.00 & y \\ \hline
9  & 29,500,5  & 1.00 & f & 0.00 & n \\ \hline
10 & 30,300,10 & 2.33 & i & 0.04 & n \\ \hline
11 & 33,380,5  & 1.63 & m & 1.00 & n \\ \hline
12 & 10,420,5  & 1.38 & m & 0.00 & n \\ \hline
13 & 25,270,5  & 2.70 & f & 0.00 & n \\ \hline
14 & 12,340,10 & 1.94 & i & 0.00 & n \\ \hline
15 & 25,300,5  & 2.33 & m & 1.00 & n \\ \hline
16 & 33,380,10 & 1.63 & m & 1.00 & n \\ \hline
18 & 21,400,10 & 1.50 & i & 0.79 & y \\ \hline
20 & 19,220,10 & 3.55 & i & 0.00 & n \\ \hline
21 & 21,400,5 & 1.50 & f & 0.00 & n \\ \hline
22 & 1,470,10  & 1.13 & i & 1.00 & n \\ \hline
24 & 33,250,5  & 3.00 & m & 0.00 & n \\ \hline
25 & 23,450,0  & 1.22 & m & 1.00 & n \\ \hline
26 & 9,300,10  & 2.33 & i & 0.00 & n \\ \hline
\end{tabular}\centering}
\subtable{
\begin{tabular}[t]{|c|c|c|c|c|c|}
\hline
ID & VELA ID & z & Type & p$_{M}$ & err \\ \hline
27 & 12,380,10 & 1.63 & i & 0.00 & n \\ \hline
28 & 6,290,5   & 2.45 & m & 1.00 & n \\ \hline
30 & 23,320,10 & 2.13 & i & 0.67 & y \\ \hline
31 & 28,330,10 & 2.03 & m & 0.00 & y \\ \hline
32 & 1,410,0   & 1.44 & m & 0.93 & y \\ \hline
33 & 5,390,10  & 1.56 & m & 0.00 & n \\ \hline
34 & 27,330,5  & 2.03 & f & 1.00 & n \\ \hline
35 & 25,270,10 & 2.70 & i & 0.00 & n \\ \hline
36 & 33,250,10 & 3.00 & m & 1.00 & n \\ \hline
37 & 29,280,10 & 2.57 & m & 1.00 & n \\ \hline
38 & 8,360,5   & 1.78 & m & 1.00 & n \\ \hline
39 & 22,460,5  & 1.17 & f & 1.00 & n \\ \hline
40 & 9,390,10  & 1.56 & m & 1.00 & n \\ \hline
41 & 29,480,5  & 1.08 & m & 1.00 & n \\ \hline
42 & 25,480,5  & 1.08 & f & 0.00 & n \\ \hline
43 & 3,380,10  & 1.63 & i & 1.00 & n \\ \hline
44 & 33,330,10 & 2.03 & m & 0.00 & y \\ \hline
46 & 17,310,5  & 2.23 & m & 1.00 & n \\ \hline
47 & 20,290,10 & 2.45 & i & 1.00 & n \\ \hline
48 & 32,320,5 & 2.12 & f & 1.00 & n \\ \hline
49 & 20,390,5  & 1.56 & m & 1.00 & n \\ \hline
50 & 24,370,5  & 1.70 & m & 1.00 & n \\ \hline
51 & 25,480,10 & 1.08 & i & 0.00 & n \\ \hline
52 & 22,460,10 & 1.17 & i & 0.00 & n \\ \hline
53 & 28,270,10 & 2.70 & m & 1.00 & n \\ \hline
\end{tabular}\centering}
\caption{Probabilities of an Individual Mock Galaxy Being in a Merger: The column ID refers to our internal mock galaxy catalogue identification number. The VELA ID column is the combination of the simulation number (sim), the time snapshot (snap), and camera id (id) as defined in \citet{simons19}. The z column corresponds to the redshift. The true type column refers to the intrinsic morphological type of the galaxy as determined by \citet{simons19}, where ''i'' refers to isolated, ''m'' refers to merger and ''f'' refers to a fake merger. A fake merger designation is for cutouts where two mock isolated galaxies where superimposed using a random separation $<$ 8'' to represent real world observations of apparent galaxy pairs that have small on-sky separations but exist at different redshifts. The merger designation includes minor-,major-, pre- post- merging/coalesced systems. The p$_M$ column refers to the probability of the object being in a merger given the classifications of 14 human classifiers and evaluated using \autoref{eq:per_gal}. The final column, err, indicates whether the probability of an object being in a merger is unconstrained where an unconstrained probability is defined as when the standard deviation of the probability distribution, p$_M$, is greater than 10\%.}
\label{tab:df_mock}
\end{center}
\end{table*}

\begin{figure*} \label{fig:isolated}
\begin{tabular}{cccc}
\subfigure[4: incorrect]{\includegraphics[width = 1.5in]{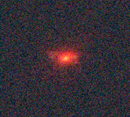}} &
\subfigure[51: correct]{\includegraphics[width = 1.5in]{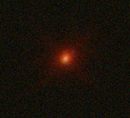}} &
\subfigure[26: correct]{\includegraphics[width = 1.5in]{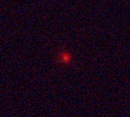}} &
\subfigure[18: incorrect]{\includegraphics[width = 1.5in]{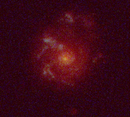}}\\
\subfigure[1: incorrect]{\includegraphics[width = 1.5in]{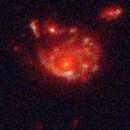}} &
\subfigure[0: correct]{\includegraphics[width = 1.5in]{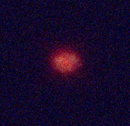}} &
\subfigure[52: correct]{\includegraphics[width = 1.5in]{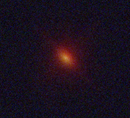}} &
\subfigure[35: correct]{\includegraphics[width = 1.5in]{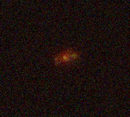}}\\
\subfigure[47: incorrect]{\includegraphics[width = 1.5in]{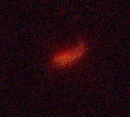}} &
\subfigure[20: correct]{\includegraphics[width = 1.5in]{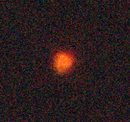}} &
\subfigure[43: incorrect]{\includegraphics[width = 1.5in]{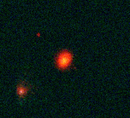}} &
\subfigure[2: correct]{\includegraphics[width = 1.5in]{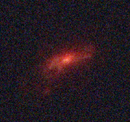}}\\
\subfigure[22: incorrect]{\includegraphics[width = 1.5in]{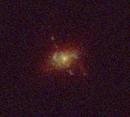}} &
\subfigure[14: correct]{\includegraphics[width = 1.5in]{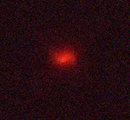}} &
\subfigure[27: correct]{\includegraphics[width = 1.5in]{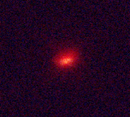}} &
\subfigure[10: correct]{\includegraphics[width = 1.5in]{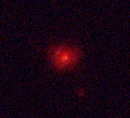}} \\
\subfigure[30: correct]{\includegraphics[width = 1.5in]{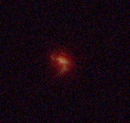}}
\end{tabular}
\caption{Isolated VELA+SUNRISE Noise Added Galaxies. The ID and measured classification are provided for each image (r:HSTWFC3/IR F160W, g:HSTACS F775W, b: ACS F435W). If a galaxy is classified correctly, p$_{M} < 1\%$, we identify it as ''correct''. All cutouts are 8''x 8''.}
\label{fig:iso}
\end{figure*}

\begin{figure*} \label{fig:fake_mergers}
\begin{tabular}{cccc}
\subfigure[3: incorrect]{\includegraphics[width = 1.5in]{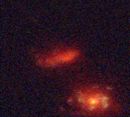}} &
\subfigure[9: correct]{\includegraphics[width = 1.5in]{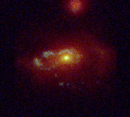}} &
\subfigure[13: correct]{\includegraphics[width = 1.5in]{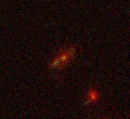}} &
\subfigure[21: correct]{\includegraphics[width = 1.5in]{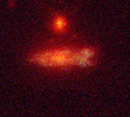}}\\
\subfigure[34: incorrect]{\includegraphics[width = 1.5in]{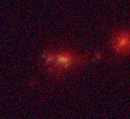}} &
\subfigure[39: incorrect]{\includegraphics[width = 1.5in]{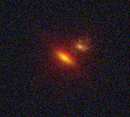}} &
\subfigure[42: correct]{\includegraphics[width = 1.5in]{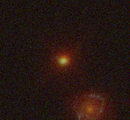}} &
\subfigure[48: incorrect]{\includegraphics[width = 1.5in]{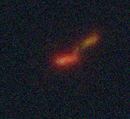}}\\
\end{tabular}
\caption{Fake Merging VELA+SUNRISE Noise Added Galaxies. The ID and measured classification are provided for each image (r:HSTWFC3/IR F160W, g:HSTACS F775W, b: ACS F435W). If a galaxy is classified correctly, p$_{M} > 99\%$, we identify it as ''correct''. All cutouts are 8''x 8''.}
\label{fig:iso}
\end{figure*}

\clearpage

\bibliography{references}{}
\bibliographystyle{aasjournal}



\end{document}